\documentclass[aps,prd,showpacs,twocolumn,superscriptaddress,nofootinbib,preprintnumbers]{revtex4-1}

\usepackage{aas_macros}

\newcommand\kmax{k_{\rm max}}

\newcommand{\La}{\text{Ly}\alpha}

\newcommand\Mpc{\text{Mpc}}
\newcommand\hMpc{\,h\text{Mpc}^{-1}}
\newcommand{\Mpch}{h^{-1}\text{Mpc}}

\newcommand{\lya}{{\rm F}}
\newcommand{\qso}{{\rm H}}
\newcommand{\cross}{{\rm X}}
\newcommand{\lyaa}{{\rm Ly\alpha}}
\newcommand{\qsoo}{{\rm LH}}
\newcommand{\sh}{{\rm shot}}
\newcommand\q{{\bf q}}

\usepackage{float}

\usepackage{amssymb, amsmath, bm, dcolumn, epsf, graphicx, latexsym, slashed} 
\usepackage[utf8]{inputenc}
\usepackage{multirow}

\usepackage{color}

\def\be{\begin{equation}}
\def\ee{\end{equation}}
\def\bea{\begin{eqnarray}}
\def\eea{\end{eqnarray}}

\usepackage{hyperref}
\usepackage{comment}
\usepackage{booktabs}
\usepackage{diagbox}

\usepackage[normalem]{ulem}

\begin{document}

\preprint{MIT-CTP/5825}

\title{
Lyman Alpha Forest -- Halo 
Cross -- Correlations in Effective Field Theory
}

\author{Anton Chudaykin}\email{anton.chudaykin@unige.ch}
\affiliation{D\'epartement de Physique Th\'eorique and Center for Astroparticle Physics,\\
Universit\'e de Gen\`eve, 24 quai Ernest  Ansermet, 1211 Gen\`eve 4, Switzerland}
\author{Mikhail M. Ivanov}
\email{ivanov99@mit.edu}
\affiliation{Center for Theoretical Physics, Massachusetts Institute of Technology, 
Cambridge, MA 02139, USA}
\affiliation{The NSF AI Institute for Artificial Intelligence and Fundamental Interactions, Cambridge, MA 02139, USA}

\begin{abstract} 
We provide a perturbative effective field theory (EFT) 
description for anisotropic (redshift-space) correlations 
between the Lyman alpha forest and a generic biased tracer of matter, 
which could be represented by quasars, high-redshift galaxies, or 
dark matter halos. 
We compute one-loop EFT power spectrum predictions for the combined analysis of the 
Lyman alpha and biased tracers' data and test them on the publicly available 
high fidelity Sherwood simulations. We use massive and light dark matter halos 
at redshift $z=2.8$ as proxies for quasars and high-redshift galaxies, respectively.
In both cases, we demonstrate that our EFT model can consistently describe the 
complete data vector consisting of the Lyman alpha forest auto spectrum, the halo auto spectrum,
and the Lyman alpha -- halo cross spectrum.
We show that the addition of cross -- correlations significantly sharpens
constraints on EFT parameters of the Lyman alpha forest and halos.
In the combined analysis, our EFT model fits the simulated cross-spectra with a percent level accuracy at 
$k_{\rm max}= 1~h$Mpc$^{-1}$,
which represents a significant improvement 
over previous analytical models. 
Thus, our work provides precision theoretical tools for
full-shape analyses of Lyman alpha -- quasar cross -- correlations 
with ongoing and upcoming spectroscopic surveys. 
\end{abstract}

\maketitle

\section{Introduction}
\label{sec:intro}

The Lyman alpha (Ly$\alpha$) forest is a collection of absorption lines 
in the spectra of distant quasars produced by neutral hydrogen clouds 
in the intergalactic medium at redshifts $2\lesssim z\lesssim 5$. 
Fluctuations in the flux transmitted through these clouds
correlate with matter distribution on cosmological scales.   
The forest thus provides
unique information about the large-scale structure 
of the Universe at high redshift, which has been extensively
used to constrain the physics of
neutrinos, dark matter, 
and dark energy~\cite{Seljak:2005, Viel:2010, PYB13, Palanque2020,Afshordi:2003, Murgia:2019,Viel:2013, Baur:2016, Irsic17, Kobayashi:2017, Armengaud:2017, Murgia:2018,Garzilli:2019, Irsic:2020, Rogers:2022, Villasenor:2023, Irsic:2023,2023PhRvL.131t1001G}. In particular, the $\La$ forest has been used 
for precision measurements of Baryon Acoustic Oscillations (BAO),
periodic fluctuations in the matter density imprinted 
in the early universe\cite{McDonald:2007,Slosar2013, Busca:2013, duMasdesBourboux:2020pck, DESI_lya_2024}. 

The cosmological information from the $\La$ forest 
can be significantly amplified by using correlations
between the forest and high redshift quasars~\cite{Font-Ribera:2013fha,
BOSS:2013igd,
duMasdesBourboux:2020pck, DESI_lya_2024}.
This includes both the BAO and the broadband shape of the 
$\La$ forest correlations
\cite{Gerardi:2022ncj, Cuceu:2021hlk}, which have recently attracted 
a significant attention as a powerful complimentary probe. 
A theoretical challenge associated with this probe is an 
accurate modeling of $\La$ forest -- quasar cross-correlations,
starting with the simplest two-point function and its Fourier transform, the power spectrum.
While on the largest scales (wavenumbers $k\lesssim 0.1~\hMpc$) the linear theory description
of the $\La$ forest -- quasar power spectrum
is adequate, it is expected to fail for larger wavenumbers,
which still carry significant
cosmological information. The non-linear $\La$ forest -- quasar correlations are challenging 
to model with hydrodynamical simulations because they require both large volumes 
and high resolution.

An inexpensive alternative to simulations is non-linear cosmological 
perturbation theory. While limited to scales $k\lesssim 1~\hMpc$
where non-linear corrections are small,
it provides a high level of accuracy and flexibility.
The framework of effective field theory (EFT) for 
large-scale structure~\cite{Baumann:2010tm,Carrasco:2012cv} 
(see~\cite{Ivanov:2022mrd} for a recent review) 
provides a systematic program of building consistent perturbation theory based only 
on symmetries and dimensional analysis.
EFT has recently become a standard tool to analyze the clustering 
of galaxies and quasars~\cite{Ivanov:2019pdj,DAmico:2019fhj,Chen:2021wdi,Chudaykin:2022nru,Chen:2024vuf,DESI:2024hhd}.
EFT for the $\La$ forest has been developed in~\cite{Ivanov:2023yla},
which was based on the formalism
of EFT for galaxies in the presence of 
selection effects~\cite{Desjacques:2018pfv}.
Previous important 
perturbation theory studies of the $\La$ forest 
include~\cite{Garny:2018byk,Garny:2020rom,Chen:2021rnb,Givans:2020sez,Givans:2022qgb}.

The main goal of this publication is to develop 
an EFT for the $\La$ forest -- quasar 
cross correlations. 
In EFT, different biased tracers of matter
are described within the same effective bias expansion, 
such that the physical differences between galaxies, 
quasars, and dark matter halos appear only
in the values of bias parameters. Therefore, the EFT description
that we lay out here is equally applicable for all these different 
tracers of matter. 
We will test our description against simulated 
Ly$\alpha$ --  halo cross-correlations, in which we use 
light and massive halos as proxies for quasars and galaxies, 
respectively. 
Our work thus can be seen as an extension of ref.~\cite{Givans:2022qgb}
that first studied the non-linear corrections to the 
Ly$\alpha$ -- halo cross spectrum. Importantly, ref.~\cite{Givans:2022qgb}
pointed out that linear theory becomes inadequate already 
on relatively large scales $\sim 0.3~\hMpc$, which motivates
the development of a systematic non-linear model
which we provide here. 

Note that in contrast with usual biased tracers, 
EFT for the $\La$ forest
requires a different type of the bias expansion
that accounts for the fact that the $\La$
forest fluctuations are only symmetric w.r.t. 
rotations around the line of sight, as opposed 
to quasars or halos that enjoy the full three-dimensional
rotational symmetry.

In addition to presenting the theory, we 
develop a full EFT-based pipeline for the combined analysis of the Ly$\alpha$ 
forest and halo auto-power spectra, as well as the Ly$\alpha$ -- halo cross-correlation.
In particular, we study the dependence of 
our result on the choices
of the covariance matrices.
Our pipeline can be readily applied to 
$\La$ forest-quasar 
data from DESI~\cite{DESI:2022,Gordon:2023dua}. 

Our work is structured as follows. We outline our methodology 
and present the simulation data in Section~\ref{sec:data}. 
There we give the details of the $\La$--halo cross spectrum
calculated in effective field theory at the one-loop order.
Our main results are summarized in Section~\ref{sec:res}. 
Section~\ref{sec:conc} draws conclusions and lists directions
for
future exploration.
In Appendix~\ref{app:lya} we present the results from the $\lyaa$ forest auto-power spectrum for various data cut choices.

\section{Data and methodology}
\label{sec:data}

\subsection{Data}
\label{sec:data1}

We use the Sherwood suite of hydrodynamic simulations~\cite{Bolton:2016bfs}.
These are large publicly available 
high-resolution simulations of the intergalactic medium with up to 17.2 billion particles. 
The fiducial cosmology of these simulations is a flat $\Lambda$CDM model with $\Omega_m=0.308$, $\Omega_b=0.0482$, $\sigma_8=0.829$, $n_s=0.961$, $h=0.678$.

In this work, we present results from the largest simulation box \texttt{L160\_N2048}, which has a box size of $L=160\,h^{-1}\Mpc$ and contains $N=2048^3$ dark matter and gas particles.
The simulations assume a homogeneous ionising background model, where the gas is in equilibrium and is optically thin~\cite{Bolton:2016bfs}.
Halo catalogs were generated with the friends of friends algorithm.
In the main text, we focus on the snapshot at the redshift $z=2.8$ which 
was previously analyzed in detail in ref.~\cite{Givans:2022qgb}.

For the halo power spectrum, we consider two different samples: a catalog containing the most massive halos with $11.5<\log_{10} (M/(h^{-1}M_\odot))<14$ and another catalog that contains all available halos.
The first catalog is aimed to simulate the clustering properties of quasars, which are typically hosted by dark matter halos heavier than $10^{12}M_\odot$~\cite{Givans:2022qgb}. 
Quasars of this type have been measured by the eBOSS collaboration and serve as a primary probe for DESI~\cite{Chudaykin:2022nru,Gordon:2023dua}.
The full halo catalog is dominated by significantly lighter halos, which act as proxies for high redshift galaxies that will be targeted by upcoming spectroscopic surveys such as Spec-S5 and WST. Indeed the $\La$ emitters have $b_1\simeq 1.5$ at $z\approx 3$~\cite{Ravi:2024twy},
very similar to the linear bias of our full halo catalog.
We denote the catalogs of the most massive halos and all halos as $\qso$ and $\qsoo$, respectively.
The halo number densities for these two catalogs at $z=2.8$ are:
\be 
\label{ng}
\begin{split}
\text{Massive halos (H)}:~&\bar{n}_h^{-1}=188.61\,h^{-3}\Mpc^3\,,\\
\text{Light halos (LH)}:~&\bar{n}_h^{-1}=0.37\,h^{-3}\Mpc^3\,.
\end{split}
\ee
The catalog of the massive halos has a lower number density, which results in a higher shot-noise value.

In this work, we utilize the 3D auto-power spectra of Ly$\alpha$ forest and halos, along with the Ly$\alpha$ -- halo cross-power spectrum from~\cite{Givans:2022qgb}.
The measurements are presented as a function of wavenumber $k$ and the cosine of the angle between the corresponding Fourier mode and the line of sight $\mu$. The $k$ space is sampled by 20 log-spaced bins in the range $[k_F , k_{\rm Ny}]$, where $k_F=2\pi/L=0.039$ is the fundamental mode of the box, and $k_{\rm Ny}$ is the Nyquist frequency. The $\mu$ is sampled by 16 uniformly spaced bins in the interval $[0,1]$.

\subsection{Methodology}
\label{sec:data11}

We perform a Markov Chain Monte-Carlo analysis to sample from the posterior distribution of EFT parameters
to access the performance of the EFT model. We adopt a Gaussian likelihood defined as
\be\begin{split}
& -2\ln \mathcal L_P = 
\left(\boldsymbol{\mathcal{P}} - \boldsymbol{\mathcal{P}}_{\mathrm{data}}\right)^{t}\cdot \left[ \boldsymbol{\mathcal{C}} \right]^{-1} \cdot \left(\boldsymbol{\mathcal{P}} - \boldsymbol{\mathcal{P}}_{\mathrm{data}}\right)
\,,
\end{split}
\ee
where $\boldsymbol{\mathcal{P}}$ and $\boldsymbol{\mathcal{P}}_{\mathrm{data}}$ represent the multi-dimensional theory and data vectors, respectively, and $\boldsymbol{\mathcal{C}}$ is the data covariance matrix.
In the most general setup, the theory vector is composed of three different spectra,
\be
\label{Pmulti}
\boldsymbol{\mathcal{P}}=(P^\lya,P^\qso,P^\cross)\,,
\ee
where the uppercase indices ``$\lya$'', ``$\qso$'' and ``$\cross$'' correspond to the Ly$\alpha$ power spectrum, halo power spectrum, and the Ly$\alpha$ -- halo cross spectrum, respectively.~\footnote{The results of this section apply to both the massive halo and all-halo catalogs, which, for simplicity, we will collectively refer to as ``$\qso$''.}
These spectra are evaluated at the grid points ($k_i$,$\mu_i$).
We assume a Gaussian covariance matrix, whose linear theory expression is given by:
\be
\label{cov}
\boldsymbol{\mathcal{C}} = \begin{pmatrix}
C^{\lya\lya} & C^{\lya\qso} & C^{\lya\cross} \\
C^{\lya\qso} & C^{\qso\qso} & C^{\qso\cross} \\ 
C^{\lya\cross} & C^{\qso\cross} & C^{\cross\cross}
\end{pmatrix}\,,
\ee
where each block represents a diagonal matrix defined as
\be\label{covL}
\begin{split}
& C^{\lya\lya}_{ii}=2N_i^{-1}P^\lya_iP^\lya_i\\
& C^{\lya\qso}_{ii}=2N_i^{-1}P^\cross_iP^\cross_i\\
& C^{\lya\cross}_{ii}=2N_i^{-1}P^\lya_iP^\cross_i\\
& C^{\qso\qso}_{ii}=2N_i^{-1}P^\qso_iP^\qso_i\\
& C^{\qso\cross}_{ii}=2N_i^{-1}P^\qso_iP^\cross_i\\
& C^{\cross\cross}_{ii}=N_i^{-1}[P^\cross_iP^\cross_i+P^\lya_iP^\qso_i]\\
\end{split}
\ee
In these expressions, $N_i$ is a number of modes in the $(k_i,\mu_i)$ bin.
This covariance structure introduces additional correlations between different spectra, which impose additional constraints on the inferred statistics.
To assess the impact of these cross-correlations, we also perform an analysis with a diagonal covariance 
obtained by neglecting the off-diagonal terms: $C^{ab}=0$ for $a\neq b$ where $a,b=\{{\rm F},{\rm H},{\rm X}\}$.
This choice represents a more conservative approach, providing more flexibility in modeling the individual spectra.
Additionally, it will allow us to validate the results obtained using the non-diagonal covariance, assessing its impact on the performance of the EFT model.

The individual power spectra $P^a$ used in the covariance can either be extracted from data or predicted using a theoretical model.
We found that the data measurements on large scales are affected by sample noise, 
which has a significant impact on the posterior distribution for bias parameters.~\footnote{
This effect is more pronounced for the catalog of the most massive halos, whereas it is less important for the total halo catalog.}
To achieve accurate covariance predictions, we employ a hybrid approach that combines the one-loop perturbation theory model with data measurements.
In practice, we apply the following algorithm.
\begin{enumerate}
\item 
We compute the theoretical spectra $P^a$ at the maximum point of the posterior in the range $k<k_{\rm max,\,fid}^a$, where $k_{\rm max,\,fid}^a$ denotes some fiducial configuration. 
For both individual and combined 3-spectra analyses, we use the values 
\[
(k_{\rm max,\,fid}^\lya,k_{\rm max,\,fid}^\qso,k_{\rm max,\,fid}^\cross)=(2,0.8,1)\hMpc~\,.
\]
\item We construct the covariance matrix based on \eqref{cov} and \eqref{covL}, using the one-loop theory predictions for $k < k_{\rm max,\,fid}^a$ and data measurements for $k > k_{\rm max,\,fid}^a$.
\item Using this covariance matrix, we perform an MCMC analysis with the fiducial data cuts.
\end{enumerate}
This process is iterated until convergence, defined as a change of less than 1\% in the bias parameter constraints.
Upon achieving convergence, the covariance matrix corresponding to the fiducial data cuts is obtained.
For arbitrary data cuts, the covariance matrix is constructed by combining the theoretical model for $k < k_{\rm max,\,fid}^a$ and data measurements for $k > k_{\rm max,\,fid}^a$.
This approach ensures an accurate covariance prediction that remains robust on large scales while maintaining accuracy on small scales.

A comment on the Gaussian approximation is in order.  
While the Gaussian diagonal covariance provides a reliable measure of the statistical error it may be inaccurate on small scales, relevant for our analysis. Previous studies~\cite{Wadekar:2019rdu,Wadekar:2020hax,Philcox:2020zyp} showed that the Gaussian covariance for the galaxy power spectrum is highly accurate on mildly non-linear scales because the analysis is effectively dominated by the theoretical error introduced by marginalization over nuisance parameters~\cite{Baldauf:2016sjb,Chudaykin:2019ock,Chudaykin:2020hbf}. 
While we expect the same argument to hold for the EFT of the $\La$ forest, 
we note that, strictly speaking, it remains an assumption
whose validation on different covariance matrices (e.g. analytic vs. empirical covariance
based on log-normal mocks) is left for future work.
Therefore, 
we proceed with the covariance choices available to us, but 
caution that the interpretation of our results is contingent on the assumptions made about the covariance matrix.

Importantly, unlike~\cite{Givans:2022qgb} we do not introduce a noise floor when modelling the covariance matrices. This allows us to perform a more stringent test of the theoretical model,
which precisely aims at extracting the information from large wavenumbers, which would be washed out by the noise floor.

\subsection{Theoretical model}
\label{sec:data2}

The theory vector $\boldsymbol{\mathcal{P}}$ is a one-loop EFT model that includes all necessary ingredients relevant on mildly non-linear scales. 
This is based on a perturbative expansion involving the most general operators that respect spacetime symmetries of the problem and the equivalence principle.
The EFT-based model for the $\lyaa$ forest auto-power spectrum was formulated in~\cite{Ivanov:2023yla} and later applied to the eBOSS 1D flux power spectrum of~\cite{Chabanier:2019}
in~\cite{Ivanov:2024jtl}. Here, we extend this approach by applying it to all three spectra: $P^\lya$, $P^\qso$, and $P^\cross$. Note that our cross-correlation model has been
recently applied to estimate the shift of the BAO peak
in the quasar-$\La$ cross spectra in~\cite{deBelsunce:2024rvv}. We provide 
the details in this publication.

The general idea of our model is that the overdensity of halos, $\delta_\qso$, and fluctuations of the $\La$
flux, $\delta_\lya$, individually can be perturbatively expanded over the linear matter density field.
Without loss of generality, the non-linear density field $\delta_a$ with $a=\{\lya,\qso\}$
can be expressed at cubic order in the linear matter overdensity $\delta^{(1)}$ as
\be
\label{delta}
\begin{split}
 \delta_a({\bf k})= &\sum_{n=1}^3 \Big[ \prod_{j=1}^n\int_{}\frac{d^3{\bf k}_j}{(2\pi)^3} \delta^{(1)}({\bf k}_j)\Big]
 K_n({\bf k}_1,...,{\bf k}_n)\\
 &\times (2\pi)^3\delta^{(3)}_D({\bf k}-{\bf k}_1-...-{\bf k}_n)\\
 &-\sum_{n=0}^2c^{a}_{2n} \mu^{2n} k^2 \delta^{(1)}({\bf k})\\
 &-\tilde{c}^a K_1({\bf k})k^4 f^4 \mu^4 \delta^{(1)}({\bf k})+\varepsilon_a({\bf k})\,,
 \end{split}
\ee
where $K_n$ represent the EFT kernels,
$c_n$ are counterterms, 
$\tilde{c}$ is the higher-order 
counterterm needed to capture
non-linear redshift space distortions 
of the collapsed tracer~\cite{Ivanov:2019pdj,Chudaykin:2020hbf,Taule:2023izt},
and 
$\varepsilon_a({\bf k}$)
is the stochastic component
uncorrelated with the linear density. 
The bias expansion is controlled by symmetries of the problem, so the $K_n$ functions are different for halos and $\lyaa$ forest.
For halos, the $K_n$ are the standard redshift-space kernels, commonly used in EFT, see e.g.~\cite{Ivanov:2019pdj}.
The $\lyaa$ forest introduce selection effects, leading to new line-of-sight dependent operators, specified in ref.~\cite{Ivanov:2023yla}.
The auto-power spectrum is defined as 
\be
\label{PkAuto}
\langle\delta_a({\bf k})\delta_a({\bf k'})\rangle=(2\pi)^3P_{\rm 1\text{-}loop}^a(k)\delta^{(3)}_D({\bf k}+{\bf k'})
\ee
where $a=\{\lya,\qso\}$.
For the cross spectrum, we write
\be
\label{PkCross}
\langle\delta_\lya({\bf k})\delta_\qso({\bf k'})\rangle=(2\pi)^3P_{\rm 1\text{-}loop}^\cross(k)\delta^{(3)}_D({\bf k}+{\bf k'})
\ee

We will use the following definition 
for the auto-power spectrum of the stochastic field
$P_{\rm stoch}$:
\be
\label{Pkstoch}
\langle\epsilon_a({\bf k})\epsilon_a({\bf k'})\rangle=(2\pi)^3P_{\rm stoch}^a(k)\delta^{(3)}_D({\bf k}+{\bf k'})
\ee
with $a=\{\lya,\qso\}$, and similarly for the cross-term $\langle\epsilon_\lya({\bf k})\epsilon_\qso({\bf k'})\rangle$,
whose spectrum we denote $P^\cross_{\rm stoch}$.

The anisotropic power spectra are calculated using the FFTLog approach
embodied in the CLASS-PT code~\cite{Chudaykin:2020aoj}. The details
can be found in ref.~\cite{Simonovic:2017mhp,Ivanov:2023yla}.
In our theory models, we use an approximate description
of the non-linear damping of the BAO signal in the linear power spectrum
by means of the isotropic 
damping factor derived in~\cite{Blas:2015qsi,Blas:2016sfa}.
Specifically, we apply the isotropic (real space) one-loop
IR resummed formula from~\cite{Blas:2016sfa}.
While it is straightforward to 
implement the full anisotropic 
suppression derived in~\cite{Ivanov:2018gjr},
this is not required for our
work given large statistical
errors of the Sherwood data
at the BAO wavenumbers $k\sim 0.1~\hMpc$.
In all expressions given below
the one-loop
IR resummation is assumed by default.  

Below, we provide explicit expressions of the $P^\lya$, $P^\qso$, and $P^\cross$ spectra individually. 
To avoid clutter, we will omit the uppercase index ``$\lya$'' for the EFT parameters and kernels associated with the $\lyaa$ forest auto-power spectrum. 

\subsubsection{Ly$\alpha$ forest auto-power spectrum}

We start with the $\lyaa$ forest auto-power spectrum.
In this case, the relevant operators are scalars under $SO(2)$ rotations around the line-of-sight.
This implies a greater flexibility in the EFT bias expansion compared to the case of galaxies, leading to the new line-of-sight dependent operators, originally derived in~\cite{Desjacques:2018pfv}.
By performing a direct calculation of eq.~\eqref{PkAuto}, we arrive at the one-loop power spectrum for $\lyaa$ forest,
\be
\label{P1loop}
\begin{split}
P_{\rm 1\text{-}loop}^\lya(k,\mu) &= K_1^2(\textbf{k})P_{\text{lin}}(k) \\
& +2\int_\q K_2^2(\q,\textbf{k}-\q)
P_{\text{lin}}(|\textbf{k}-\q|)P_{\text{lin}}(q)  \\
&+ 6 K_1(\textbf{k})P_{\text{lin}}(k)\int_\q K_3(\textbf{k},-\q,\q)P_{\text{lin}}(q)\\
&-2(c_0+c_2\mu^2+c_4\mu^4)K_1(\textbf{k})k^2P_{\text{lin}}(k)\,.
\end{split}
\ee
The linear EFT kernel is expressed as
\be
\label{K1}
K_1=b_1-b_\eta f\mu^2
\ee
where $b_1$ is the selection-free linear bias and $b_\eta$ is the new selection-dependent bias parameter.
The $K_2$ and
$K_3$ introduce new selection-dependent EFT operators
which are absent in the case of galaxies.
Explicit expressions for these non-linear operators are provided in ref.~\cite{Ivanov:2023yla}.
Next, the $k^2P_{\text{lin}}$ corrections accounts for the higher-derivative contributions.
The main role of these terms is to absorb the UV dependence of the loop integrals.
Although these nominally contribute at the three-loop order and could be ignored at the 1-loop level, the inclusion of these parameters noticeably improves the fit~\cite{Ivanov:2023yla}, so we opted to retain $c_{i}$.
Finally, we neglect the stochastic contributions. The 
physical stochastic contributions 
are strongly suppressed due to high column densities of the $\lyaa$ forest~\cite{McQuinn:2011pt}.
Note that in EFT 
non-zero stochasticity
parameters are expected to be 
generated by the UV parts of loop
integrals, but their effect
is of the order of two-loop
corrections at $z\approx 3$,
see~\cite{deBelsunce:2024rvv}
for a detailed discussion
and explicit tests.

The 1-loop EFT model for $P^\lya$ depends on 16 free parameters: 2 linear biases, 11 non-linear biases and 3 higher-derivative operators.  
We impose the following priors on these parameters,
\be
\label{lyaparams}
\begin{split}
& b_1\in [-2,2]\,,\quad  b_\eta\in [-2,2]\,,\quad b_2\sim \mathcal{N}(0,2^2)\,,\\
&b_{\mathcal{G}_2}\sim \mathcal{N}(0,2^2)\,,\quad b_{\Gamma_3}\sim \mathcal{N}(0,1^2)\\
& b_{\eta^2} \sim \mathcal{N}(0,2^2)\,,\quad  b_{\delta \eta} \sim \mathcal{N}(0,2^2)\,,\\
& b_{(KK)_\parallel} \sim \mathcal{N}(0,2^2)\,,\quad 
b_{\Pi^{[2]}_\parallel} \sim \mathcal{N}(0,2^2)\,,\\
& b_{\Pi^{[3]}_\parallel} \sim \mathcal{N}(0,2^2)\,,\quad b_{\delta\Pi^{[2]}_\parallel} \sim \mathcal{N}(0,2^2)\,,\\
& b_{(K\Pi^{[2]})_\parallel} \sim \mathcal{N}(0,2^2)\,,\quad 
b_{\eta\Pi^{[2]}_\parallel} \sim \mathcal{N}(0,2^2)\,,\\
&\frac{c_{0,2,4}}{[\Mpch]^2} \sim \mathcal{N}(0,1^2)\,,
\\
\end{split} 
\ee
where $\mathcal{N}(\mu,\sigma^2)$ stands for a Gaussian distribution 
with mean $\mu$ and r.m.s. $\sigma$.
Our analysis differs from the previous work~\cite{Ivanov:2023yla} in several aspects. 
First, we include the cubic bias $b_{\Gamma_3}$ because it affects the
parameter error bars in the combined analyses.
Second, we broaden the priors on the $c_{0,2,4}$, as we use a slightly different convention~\eqref{P1loop}.
Jumping ahead, let us note that we will find that the values of the higher-derivative parameters are of order $10^{-2}$ with errorbars that are much tighter than the priors. These values 
are consistent with the naturalness arguments
that $c_i\sim k_{\rm NL}^{-2}$.

\subsubsection{Halo auto-power spectrum}

For the the halo power spectrum, we exploit the standard EFT model from~\cite{Chudaykin:2020ghx,Chudaykin:2024wlw}.
This is based on a larger $SO(3)$ symmetry, which reduces a number of possible operators compared to the Lyman alpha forest case.
In the absence of selection effects, the linear kernel takes the standard form,
\be
\label{K1}
K_1^\qso=b_1^\qso+f\mu^2
\ee
where $b_1^\qso$ denotes the linear halo bias.

The $P^\qso$ model includes the stochastic contribution, which at 1-loop order is given by 
\be
\label{Pstoch}
P_{\rm stoch}=\frac{1}{\bar n_h}\left[1+
P_{\rm shot}^\qso
+ a_0^\qso \left(
\frac{k}{k_{\rm NL}}\right)^2 + a_2^\qso\mu^2 \left(
\frac{k}{k_{\rm NL}}\right)^2 
\right]\,,
\ee
where $P_{\rm shot}^\qso$ is residual constant shot-noise contribution, $a_0^\qso$ and $a_2^\qso$ are scale-dependent stochastic biases. We define the non-linear scale as $k_{\rm NL}=3\,h\Mpc^{-1}$
following~\cite{Ivanov:2023yla}.
It should be noted that the constant shot-noise $\bar n_h^{-1}$ is subtracted from the halo auto-power spectrum data. The stochastic EFT parameters are expected to be $\mathcal{O}(1)$ numbers.

We also include a higher-order derivative correction, as the halos are virialized objects with significant velocities.
Following~\cite{Ivanov:2019pdj}, we incorporate the $k^4$ redshift-space counterterm,
\be
\label{Pctr2}
P_{\rm \nabla^4_z\delta}(k,\mu)=-\tilde{c}^\qso f^4\mu^4k^4\big[K_1^\qso\big]^2P_{\text{lin}}(k)
\ee
where $\tilde c^\qso$ is the higher-order counterterm. 

The 1-loop EFT model for $P^\qso$ features 11 EFT parameters.
We adopt the following priors,
\be
\label{eq:fitparams}
\begin{split}
& b^\qso_1\in [0,10]\,,\quad  b^\qso_2\sim \mathcal{N}(0,2^2)\,,\\
&b^\qso_{\mathcal{G}_2}\sim \mathcal{N}(0,2^2)\,,\quad b^\qso_{\Gamma_3}\sim \mathcal{N}(0,2^2)\\
&\frac{c^\qso_{0,2,4}}{[\Mpch]^2} \sim \mathcal{N}(0,10^2)\,,\quad \frac{\tilde{c}^\qso}{[\Mpch]^4} \sim \mathcal{N}(0,100^2)\\
&P_{\rm shot}^\qso \sim \mathcal{N}(0,1^2)\,,\\
& a_{0}^\qso \sim \mathcal{N}(0,1^2)\,,
\quad a_2^\qso\sim \mathcal{N}(0,1^2)\,,
\end{split} 
\ee
For $b_1^\qso$, $b_2^\qso$, $b_{\mathcal{G}_2}^\qso$ and $b_{\Gamma_3}^\qso$ we adopt the same priors as in the $P^\lya$ case~\eqref{lyaparams}.
For the higher-derivative contributions, we choose large uninformative
priors.
The priors on stochastic bias parameters are motivated by the EFT naturalness arguments in the physical units~\cite{Chudaykin:2020aoj}.

A comment is in order on the $c_4$ parameter that appears in front of the $k^2\mu^4 P_{\rm lin}$
counterterm. For tracers without selection effects, this counterterm must be universal as 
dictated by the equivalence principle~\cite{Perko:2016puo,Ivanov:2024xgb}.
For the Lyman $\alpha$ forest fluctuations, however, there is a higher derivative line-of-sight
counterterm present at the level of the bias expansion,
\be 
\delta_F|_{\rm higher-deriv}\supset \hat{z}^i \hat{z}^j\hat{z}^k \hat{z}^l\partial_i\partial_j\partial_k\partial_l \Phi\,,
\ee
where $\Phi$ is Newton's potential and $\hat z^i$ is the unit line-of-sigh 
vector. 
This leads to $c_4\neq c_4^H$
for the combined 3-spectra analysis with $P^\qso$.
This implies that $P^\qso$ introduces 11 new EFT parameters.

\subsubsection{Ly$\alpha$ -- halo cross-power spectrum}

The computation of the Ly$\alpha$ -- halo cross-power spectra \eqref{PkCross} involves symmetrizing the expression with respect to the $\lyaa$ and halo tracers. 
The direct calculation lead us to the following expression,
\be
\label{P1loopX}
\begin{split}
&P^\cross_{\rm 1\text{-}loop}(k,\mu) = K_1(\textbf{k})K_1^\qso(\textbf{k})P_{\text{lin}}(k) \\
& +2\int_\q K_2(\q,\textbf{k}-\q)K_2^\qso(\q,\textbf{k}-\q)
P_{\text{lin}}(|\textbf{k}-\q|)P_{\text{lin}}(q)  \\
&+ 3 P_{\text{lin}}(k)\!\!\int_\q [K_1(\textbf{k}) K_3^\qso(\textbf{k},-\q,\q) \\
&+ K_1^\qso(\textbf{k}) K_3(\textbf{k},-\q,\q)]P_{\text{lin}}(q)\\
&-(c_0+c_2\mu^2+c_4\mu^4)K_1^\qso(\textbf{k})k^2P_{\text{lin}}(k)\\
&-(c^\qso_0+c^\qso_2\mu^2+c_4^\qso\mu^4)K_1(\textbf{k})k^2P_{\text{lin}}(k)\\
& + P^\cross_{\rm \nabla^4_z\delta} \,.
\end{split}
\ee
A comment on the stochastic 
contribution is in order. 
In galaxy muti-tracer analysis 
it is often assumed that 
two types of galaxies have zero 
stochastic cross-correlation, see e.g.
\cite{eBOSS:2020rpt}.
However, this assumption is hard to justify
from the EFT point of view. 
For two tracers A and B one could 
in general 
write down~\cite{Mergulhao:2021kip}
\be 
\begin{split}
&P_{\rm shot}^\cross (k)=\frac{1}{\sqrt{\bar n_A \bar n_B}}+O(k^2/k_{\rm NL}^2)~\,.
\end{split}
\ee
In our case effectively 
$\bar n^{-1}_{\La}\simeq 0$~\cite{McQuinn:2011pt}, plus 
the number density of halos 
is quite high even for the massive ones, 
which makes $P_{\rm shot}^\cross$ highly suppressed.

In EFT, however, $P_{\rm shot}^\cross$ also 
contains the counterterm part needed 
to cancel the UV-dependence of the 
$\langle \delta^2 \delta^2 \rangle$ - type
loop 
integrals. If we assume that the quadratic 
bias parameters are $O(1)$, this will give 
us the estimate~\cite{deBelsunce:2024rvv}: 
\be 
\label{PstochX2}
P_{\rm shot}^\cross\sim \frac{1}{k_{\rm NL}^3}\sim 0.05~[\Mpch]^3\,.
\ee 
This will generate a non-zero $P_{\rm shot}^\cross$,
but its amplitude is very small given the
error bars of the Sherwood simulation. 
Thus, we will proceed with 
$P_{\rm shot}^\cross=0$
in our main analyses, and 
use a model with non-zero 
$P_{\rm shot}^\cross$ only to
test the validity of our baseline analysis.
Jumping ahead, let us 
say here that the addition of 
$P_{\rm shot}^\cross$ does not improve 
the fit to the Sherwood data, 
and hence it is reasonable to set it to zero
following our estimates.
We note however, that for other tracers
and experiments, e.g.
eBOSS quasars with large shot noise~\cite{Chudaykin:2022nru}, 
one may need to 
include $P_{\rm shot}^\cross$ in the fit.

We introduce only one unique operator for the cross-power spectrum, the next-to-leading order $k^4$ redshift-space counterterm.
For this term, we employ a symmetrized version of~\eqref{Pctr2},
\be
\label{Pctr2cross}
P^\cross_{\rm \nabla^4_z\delta}(k,\mu)=-\tilde{c}^\cross f^4\mu^4k^4K_1(\textbf{k})K_1^\qso(\textbf{k})P_{\text{lin}}(k)
\ee
where $\tilde c^\cross$ is a free EFT parameter.
We ignore the stochastic contributions as they are expected to be negligible for the cross-power spectrum.

At face value, the 1-loop EFT model for $P^\cross$ depends on 28 free parameters: 27 terms shared with the $P^\lya$ and $P^\qso$ models, and one unique FoG operator specific to the $P^\cross$ spectrum.
We impose the broad uninformative prior on the latter parameter,
\be
\frac{\tilde{c}^\cross}{[\Mpch]^4} \sim \mathcal{N}(0,100^2)\,.
\ee

\subsection{Analysis pipeline}

We fit the multidimensional vector $\boldsymbol{\mathcal{P}}$~\eqref{Pmulti} using the 1-loop EFT model.
The theoretical calculation are carried out with a custom script interfaced with the \texttt{CLASS-PT} code~\cite{Chudaykin:2020aoj}. To sample the posterior distributions, we employ a Markov Chain Monte Carlo (MCMC) analysis.

We fix the cosmological parameters and vary only the EFT parameters. 
For example, in the combined 3-spectra analysis, we vary 28 nuisance parameters : 
\begin{widetext}
\begin{equation*}
\label{biasP}
\begin{split}
&\{b_1,b_\eta,b_2,b_{\mathcal{G}_2},b_{\eta^2},b_{\delta \eta},b_{(KK)_\parallel},b_{\Pi^{[2]}_\parallel}\big|b_{\Gamma_3},c_0,c_2,c_4,b_{\Pi^{[3]}_\parallel},b_{(K\Pi^{[2]})_\parallel},b_{\delta\Pi^{[2]}_\parallel},b_{\eta\Pi^{[2]}_\parallel}\}\\
&\times\{b_1^\qso,b_2^\qso,b^\qso_{\mathcal{G}_2}\big|b^\qso_{\Gamma_3},c^\qso_0,c^\qso_2,c^\qso_4,\tilde{c}^\qso,P^\qso_{\rm shot},a^\qso_0,a^\qso_2\}\times\{\tilde{c}^\cross\}
\end{split}
\end{equation*}
\end{widetext}
The parameters on the left side of the vertical line are directly sampled in our MCMC chains, while the parameters on the right, which appear quadratically in the likelihood, are marginalized over analytically, with their posteriors later recovered from the chains~\textit{a posteriori}.

The MCMC chains are run using the \texttt{Montepython} sampler~\cite{Audren:2012wb,Brinckmann:2018cvx}.
The plots and marginalized constraints are generated with the \texttt{getdist} package~\cite{Lewis:2019xzd}.~\footnote{\href{https://getdist.readthedocs.io/en/latest/}{https://getdist.readthedocs.io/en/latest/}}

\section{Results}
\label{sec:res}

In this section, we present our results.
We perform the analysis with both catalogs of massive and light halos.

\subsection{Massive halos}
\label{sec:res1}

\subsubsection{$P^\qso$ analysis}
\label{sec:res1_1}

We begin by analyzing the auto-power spectrum of most massive halos. 

\begin{figure}[!t]
    \centering
    \includegraphics[width=0.5\textwidth]{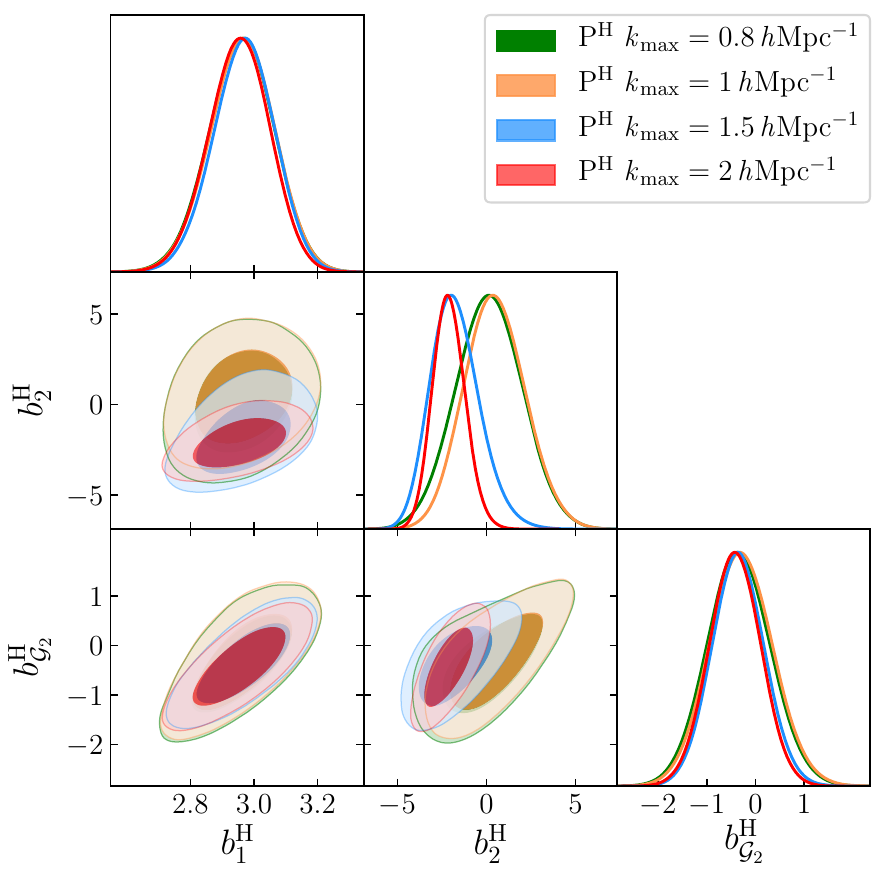}\\
    \caption{Marginalized posteriors obtained from the halo power spectrum (massive halos) for different values of $k_{\rm max}$: $0.8$, $1$, $1.5$ and $2\hMpc$ (green, orange, blue, red, respectively).}
    \label{fig:halo}
\end{figure}

Fig.~\ref{fig:halo} shows 1D and 2D marginalized posterior distributions for the bias parameters in the 1-loop EFT model.
The results are presented for four different $\kmax$ values: $0.8$, $1$, $1.5$, and $2\hMpc$.
We see that the posteriors are consistent with each other.
As a frequentest confirmation of our results, we found an equally good fit across all configurations.
Starting from $\kmax=1.5\hMpc$, we observe that the mean value of $b_2^\qso$ is shifted lower.
For $\kmax=2\hMpc$, the errors decrease further, and $b_2^\qso$ is lower than zero by $2\sigma$.
These $b_2^\qso$ values appear to be in conflict with the halo bias {prediction based on the background-split argument} from~\cite{Lazeyras:2015lgp}, $\tilde b_2^\qso=2.2$~\footnote{\label{fn:split}To estimate $\tilde b_2^\qso$, we used the best-fit values of $b_1^\qso$ and $b^\qso_{\mathcal{G}_2}$ from the baseline 3-spectra analysis, as described in Sec.~\ref{sec:comb_HM}. Note that in our convention $b_2=b_2^{\rm ref}+\frac{4}{3}b_{\mathcal{G}_2}$, where $b_2^{\rm ref}$ is the value used in ref.~\cite{Lazeyras:2015lgp}.}.
The observed shifts indicate a mild systematic bias for $\kmax>1\hMpc$, likely due to higher-order corrections not accounted for in our one-loop model.

\begin{figure}[!t]
    \centering
    \includegraphics[width=0.45\textwidth]{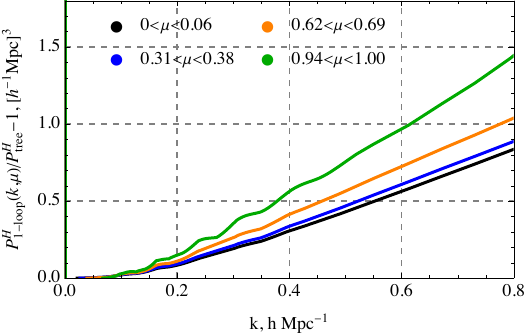}
    \caption{The magnitude of one-loop corrections relative to the linear theory prediction for massive halos. The theory prediction is based on the best-fit model with $\kmax=0.8\hMpc$.}
    \label{fig:loop}
\end{figure}

To determine a baseline $\kmax$ configuration, we quantify the magnitude of the one-loop correction as a function of wavenumber. In Fig.~\ref{fig:loop}, we plot the one-loop contribution
divided by the tree-level model.
We see that the magnitude of perturbative correction exceeds the linear theory prediction already at $\kmax=0.8\hMpc$ for $\mu\gtrsim0.6$, suggesting that higher-loop corrections may not be negligible.~\footnote{On the practical side, it might be more optimal to employ a $\mu$-dependent $\kmax$ cutoff. We plan to explore this option in future.}
Notably, the impact of the one-loop correction increases towards $\mu = 1$.
This suggests that the velocity field is primarily responsible
for the breakdown of perturbation theory at small scales.
Indeed,  the velocity field is more nonlinear than the density field for dark matter halos. 
Strictly speaking, the 1-loop EFT is valid until $\kmax\simeq0.6\hMpc$ for the transverse ($\mu\sim0$) modes and until $\kmax\simeq0.4\hMpc$ for the modes along the line of sight ($\mu\sim1$).
Given that no biases are observed at the level of the parameter estimation, it is likely that the two-loop corrections were partly absorbed by the one-loop nuisance parameters and counterterms.
We however note that the EFT model can be still applied on these scales as a phenomenological model that is capable to describe the data with high accuracy.
Thus, we select $\kmax=0.8\hMpc$ as a baseline for the $P^\qso$ analysis.

\subsubsection{$P^\lya+P^\qso+P^\cross$ analysis: diagonal covariance}
\label{sec:comb_HM}

We present the parameter constraints from the combined analysis of the $P^\qso$, $P^\lya$ and $P^\cross$ spectra. 
We employ the Gaussian covariance matrix, neglecting the off-diagonal terms.
For the halo auto-power spectrum, we fix $\kmax^\qso=0.8\hMpc$, as validated in the previous section. For the $\lyaa$ forest auto-power spectra, we adopt $\kmax^\lya=2\hMpc$ which is lower than the $\kmax^\lya=3\hMpc$ value used in ref.~\cite{Ivanov:2023yla}.
We found that the parameter constraints derived from $P^\lya$-only analysis with $\kmax^\lya=3\hMpc$ are inconsistent with those obtained from the combined $P^\lya+P^\qso+P^\cross$ analysis.
As detailed in App.~\ref{app:lya}, this discrepancy can be attributed to higher-order corrections in the $\lyaa$ forest power spectrum, which shift the posteriors to a new minimum.
Therefore, we select a more conservative $\kmax^\lya$ value.

\begin{figure*}
    \centering
    \includegraphics[width=0.99\textwidth]{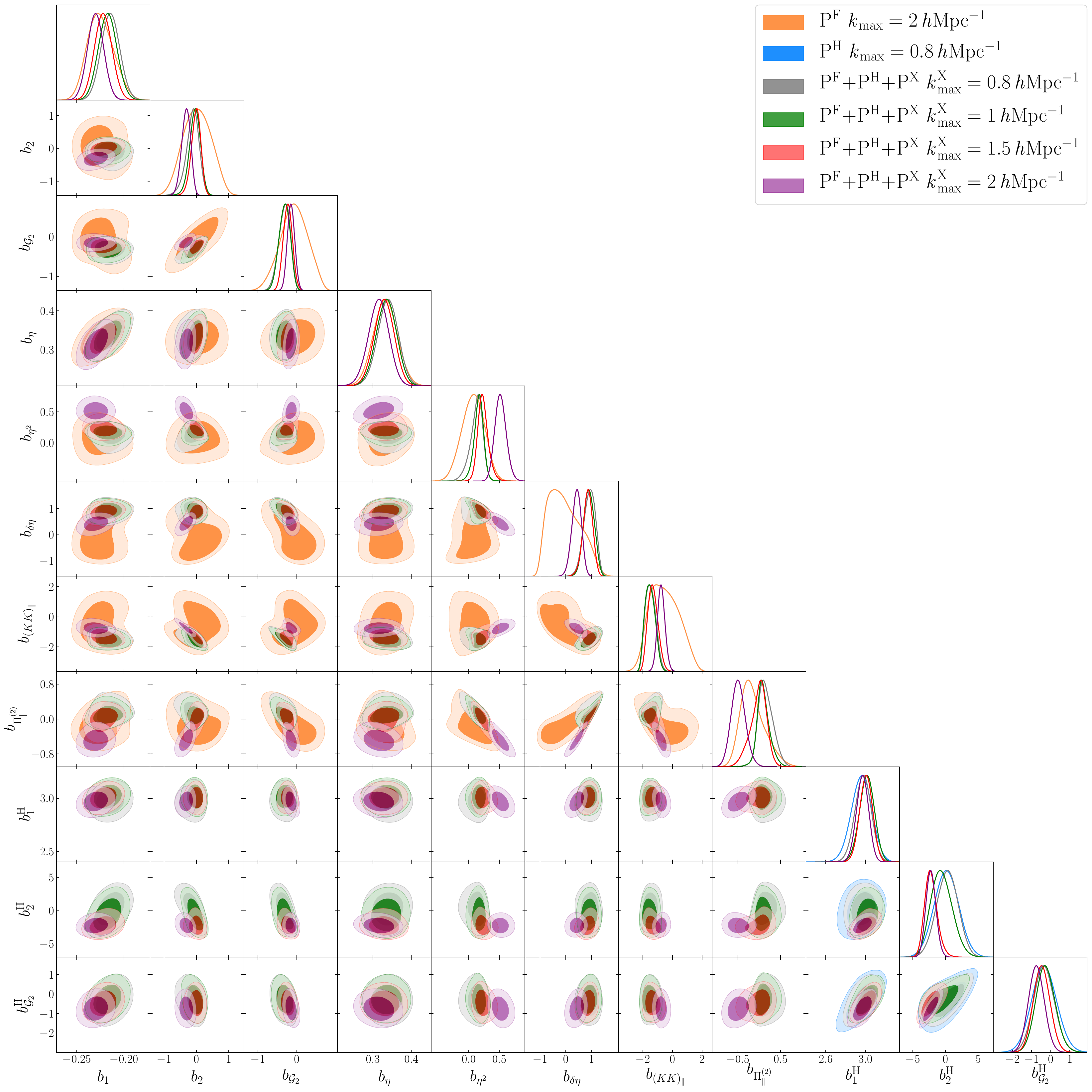}\\
    \caption{Marginalized posteriors on nuisance parameters of the EFT model for the Ly$\alpha$ forest auto-power spectrum $P^\lya$ (orange), the massive halo auto-power spectrum $P^\qso$ (blue), and their combination with the Ly$\alpha$ -- halo cross-power spectrum $P^\cross$ at $z=2.8$. The combined analysis results are shown for four different values of $k^\cross_{\rm max}$: $0.8$, $1$, $1.5$, and $2\,\hMpc$ (gray, green, red, purple, respectively). All results are obtained with the diagonal covariance.}
    \label{fig:all}
\end{figure*}

Fig.~\ref{fig:all} shows the 1D and 2D marginalized posterior distributions for the bias parameters derived from various data combinations.
The results of 3-spectra analyses are presented for four different $\kmax^\cross$ values: $0.8$, $1$, $1.5$, and $2\hMpc$.
We see that the posteriors obtained from the auto-power spectra $P^\lya$ and $P^\qso$ are fully consistent with those from the $P^\lya+P^\qso+P^\cross$ analysis up to $\kmax^\cross=1.5\hMpc$.
For $\kmax^\cross=2\hMpc$, the contours for the non-linear biases $b_{\eta^2}$ and $b_{\Pi^{[2]}_\parallel}$ are significantly shifted compared to the results with lower $\kmax^\cross$ values.
These shifts suggest that the fit is biased for $\kmax^\cross=2\hMpc$.

As a validation of our scale cuts, we perform a $\chi^2$ test for the $P^\lya+P^\qso+P^\cross$ analysis. 
For $\kmax^\cross=1\hMpc$, the nominal $\chi^2$ statistics across the 355 data points is $362$.
This indicates a good fit for 28 free parameters.
It is important to note, however, our EFT parameters are quite degenerate and hence the counting of degrees of freedom is not straightforward.~\footnote{For example, we found that adding the $b_{\Gamma_3}$ does not affect the quality of the fit.} 
The fit quality deteriorates at higher $\kmax^\cross$ values: for $\kmax^\cross = 1.5$, the $\chi^2$ values increase to $385$ for $371$ data points; for $\kmax^\cross = 2$ it rises to $441$ for $387$ data points.
These results suggest to choose $\kmax^\cross=1\hMpc$ as a baseline.

Tab.~\ref{tab:all} presents the 1D marginalized parameter constraints for the baseline analyses. 
\begin{table*}[!htb]
\renewcommand{\arraystretch}{1.2}
%\normalsize
\small
\centering
\begin{tabular}{|l||c|c|c c|}
 \hline
\multirow{3}{*}{\diagbox{ {\small Param.}}{\small Data}}
& \multirow{3}{*}{$P^\lya$}
& \multirow{3}{*}{$P^\qso$} 
& \multicolumn{2}{c|}{\multirow{2}{*}{$P^\lya+P^\qso+P^\cross$}}  
%& $P^\lya+P^\qso+P^\cross$ 
\\ 
& & & & 
\\
%\cline{4-5}
& & & \text{diag cov} & \text{off-diag cov}
\\
\hline\hline
%lya
$b_{1 }$ 
& $-0.2245_{-0.0154}^{+0.0126}$ %
& --
& $-0.2171_{-0.0102}^{+0.0099}$ %%
& $-0.2188_{-0.0092}^{+0.0089}$ %% 
\\
$b_\eta$ 
& $0.332_{-0.031}^{-0.031}$ %
& --
& $0.335_{-0.027}^{+0.028}$ %% 
& $0.332_{-0.021}^{+0.021}$ %% 
\\
$b_{2 }$ 
& $0.03_{-0.46}^{+0.44}$ %
& --
& $-0.05_{-0.16}^{+0.19}$ %% 
& $-0.08_{-0.11}^{+0.12}$ %%
\\
$b_{\mathcal{G}_2 }$ 
& $-0.07_{-0.34}^{+0.38}$ %
& --
& $-0.32_{-0.13}^{+0.16}$ %% 
& $-0.35_{-0.12}^{+0.12}$ %% 
\\
$b_{\eta^2}$  
& $0.072_{-0.180}^{+0.194}$ %
& --
& $0.163_{-0.071}^{+0.087}$ %%
& $0.198_{-0.056}^{+0.066}$ %%
\\
$b_{\delta\eta}$  
& $-0.03_{-0.83}^{+0.42}$ %
& --
& $0.90_{-0.22}^{+0.21}$ %% 
& $0.84_{-0.19}^{+0.16}$ %%
\\
$b_{(KK)_\parallel}$ 
& $-0.51_{-1.18}^{+0.86}$ %
& --
& $-1.51_{-0.38}^{+0.31}$ %% 
& $-1.44_{-0.32}^{+0.24}$ %%
\\
$b_{\Pi^{[2]}_\parallel}$  
& $-0.142_{-0.350}^{+0.212}$ %
& --
& $0.101_{-0.165}^{+0.121}$ %%
& $0.042_{-0.120}^{+0.076}$ %%
\\
\hline
$b_{\Gamma_3}$ 
& $-0.49_{-0.13}^{+0.13}$ %
& --
& $0.03_{-0.11}^{+0.11}$ %% 
& $-0.13_{-0.11}^{+0.11}$ %%
\\
$10^2 c_0/[\Mpch]^2$ 
& $-2.32_{-1.06}^{+1.06}$  %
& --
& $-2.92_{-0.27}^{+0.27}$ %%
& $-2.95_{-0.27}^{+0.27}$ %%
\\
$10^2 c_2/[\Mpch]^2$ 
& $4.26_{-1.54}^{+1.54}$ %
& --
& $5.41_{-1.06}^{+1.06}$ %%
& $4.99_{-1.05}^{+1.05}$ %%
\\
$10^2 c_4/[\Mpch]^2$ 
& $-5.38_{-1.01}^{+1.01}$ %
& -- 
& $-7.72_{-0.88}^{+0.88}$ %%
& $-7.26_{-0.88}^{+0.88}$ %%
\\
$b_{\Pi^{[3]}_\parallel}$ 
& $0.771_{-0.089}^{+0.089}$ %
& --
& $2.437_{-0.089}^{+0.089}$ %%
& $2.323_{-0.088}^{+0.088}$ %%
\\
$b_{\delta\Pi^{[2]}_\parallel}$ 
& $-0.05_{-0.19}^{+0.19}$ %
& -- 
& $0.55_{-0.19}^{+0.19}$ %%
& $0.04_{-0.19}^{+0.19}$ %%
\\
$b_{(K\Pi^{[2]})_\parallel}$ 
& $-1.64_{-0.25}^{+0.25}$ %
& --
& $-0.59_{-0.21}^{+0.21}$ %%
& $-0.53_{-0.21}^{+0.21}$ %%
\\
$b_{\eta\Pi^{[2]}_\parallel}$ 
& $-0.24_{-0.43}^{+0.43}$ % 
& --
& $-0.55_{-0.44}^{+0.44}$ %%
& $-1.69_{-0.44}^{+0.44}$ %%
\\
\hline\hline
%qso
$b_{1 }^\qso$ 
& -- 
& $2.960_{-0.102}^{+0.108}$ %
& $3.014_{-0.075}^{+0.078}$ %%
& $2.889_{-0.072}^{+0.075}$ %%
\\
$b_{2 }^\qso$ 
& -- 
& $0.16_{-1.92}^{+1.91}$ %
& $-0.55_{-1.81}^{+1.50}$ %% 
& $-0.85_{-1.09}^{+1.04}$ %%
\\
$b_{\mathcal{G}_2 }^\qso$ 
& --
& $-0.34_{-0.68}^{+0.66}$ %  
& $-0.26_{-0.54}^{+0.49}$ %% 
& $-0.83_{-0.43}^{+0.43}$ %% 
\\
\hline
$b_{\Gamma_3}^\qso$ 
& --
& $-0.03_{-0.53}^{+0.53}$ %
& $-0.11_{-0.27}^{+0.27}$ %% 
& $0.23_{-0.22}^{+0.22}$ %%
\\
$c_0^\qso/[\Mpch]^2$ 
& --
& $-0.44_{-2.94}^{+2.94}$ %
& $-0.42_{-0.12}^{+0.12}$ %%
& $0.03_{-0.11}^{+0.11}$ %%
\\
$c_2^\qso/[\Mpch]^2$ 
& --
& $2.06_{-1.11}^{+1.11}$ %
& $1.54_{-0.27}^{+0.27}$ %%
& $1.73_{-0.23}^{+0.23}$ %% 
\\
$c_4^\qso/[\Mpch]^2$ 
& --
& $3.70_{-1.33}^{+1.33}$ %
& $2.77_{-0.58}^{+0.58}$ %%
& $2.91_{-0.44}^{+0.44}$ %%
\\
$\tilde{c}^\qso/[\Mpch]^4$ 
& --
& $-1.67_{-1.42}^{+1.42}$ %
& $-0.59_{-0.71}^{+0.71}$ %%
& $-0.84_{-0.36}^{+0.36}$ %%
\\
$P_\sh^\qso$ 
& --
& $-0.066_{-0.764}^{+0.764}$ %
& $0.111_{-0.051}^{+0.051}$ %% 
& $0.084_{-0.033}^{+0.033}$ %%
\\
$a_0^\qso$ 
& --
& $0.04_{-0.97}^{+0.97}$ %
& $-0.10_{-0.90}^{+0.90}$ %%
& $0.09_{-0.72}^{+0.72}$ %%
\\
$a_2^\qso$ 
& --
& $-0.03_{-1.0}^{+1.0}$ %
& $0.03_{-0.97}^{+0.97}$ %% 
& $0.28_{-0.93}^{+0.93}$ %%
\\
\hline\hline
%cross
$\tilde{c}^\cross/[\Mpch]^4$ 
& --
& --
& $-0.27_{-0.16}^{+0.16}$ %%
& $-0.32_{-0.12}^{+0.12}$ %%
\\
 \hline
 \end{tabular} 
 \caption{One-dimensional marginalized constraints on 
 nuisance parameters of the one-loop EFT model from the Ly$\alpha$ forest auto-power spectrum with $k^\lya_{\rm max}=2\,\hMpc$ (second column), the massive halo auto-power spectrum with $k^\qso_{\rm max}=0.8\,\hMpc$ (third column) and their combination with the Ly$\alpha$ -- halo cross-power spectrum with $k^\cross_{\rm max}=1\,\hMpc$, using diagonal covariance (fourth column) and off-diagonal covariance (fifth column), at $z=2.8$. Parameter constraints related to each respective spectrum are grouped together. The parameters in the upper section were directly sampled in our MCMC chains, while the parameters in the lower section were analytically marginalized in the likelihood, with their posteriors recovered from the chains~\textit{a posteriori}.
 \label{tab:all}}
\end{table*}
First, we see that the $P^\lya$-only analysis provides the informative constraints on all Lyman alpha bias parameters within their priors.
This validates our choice of EFT priors and showcases the power of the EFT approach.
Although the leading-order counterterms for $\lyaa$ forest are of order $10^{-2}$, $c_2$ and $c_4$ are detected with high significance that motivates their inclusion in the analysis.
A second important observation is that the inclusion of $P^\cross$ significantly improves the constraints obtained from the individual auto-power spectra.
In particular, the errors on all non-linear Ly$\alpha$ forest bias parameters are reduced by more than a factor of 2, with constraints on $b_{\delta\eta}$ and $b_{(KK)_\parallel}$ improving by nearly a factor of 3 compared to the $P^\lya$-only analysis.
Additionally, the uncertainty on $c_0$ is substantially reduced, while the improvement for $c_2$, $c_4$ is more moderate. 
Surprisingly, the uncertainties on the cubic bias parameters remain largely unchanged.
The improvement for the halo bias parameters is more modest, with the most significant impact on $b_{\Gamma_3}^\qso$, whose error is reduced by a factor of 2.
At the same time, the halo counterterms and constant shot-noise parameter constraints improve dramatically, ranging from 2 to 25 times better than in the $P^\qso$-only analysis.
We conclude that the $P^\lya$, $P^\qso$, and $P^\cross$ probes are highly complementary, and their combination yields a significant information gain.

\begin{figure*}
    \centering
    \includegraphics[width=0.41\textwidth]{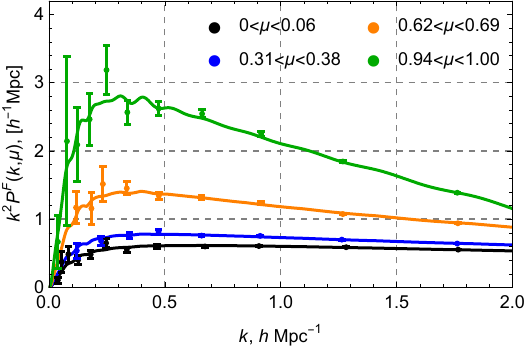}
    \includegraphics[width=0.41\textwidth]{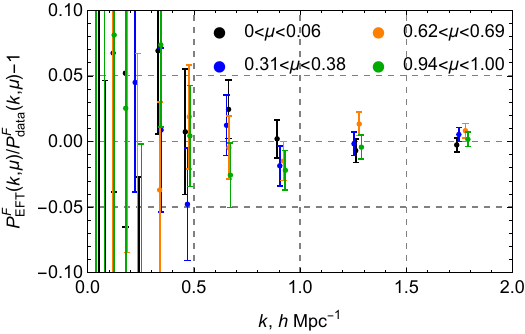}\\
    \includegraphics[width=0.41\textwidth]{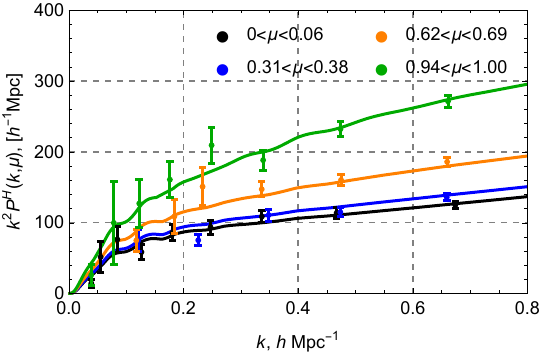}
    \includegraphics[width=0.41\textwidth]{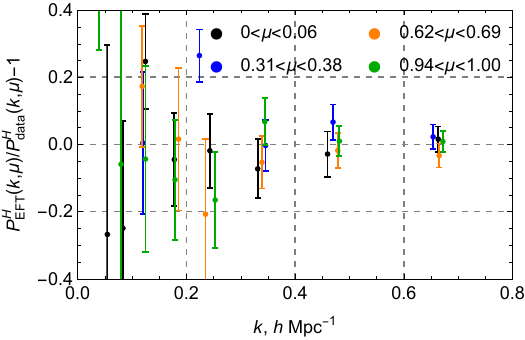}\\
    \includegraphics[width=0.41\textwidth]{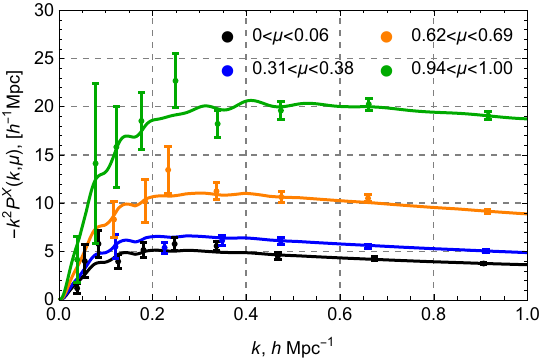}
    \includegraphics[width=0.41\textwidth]{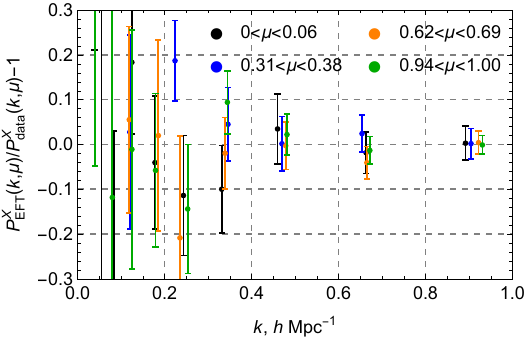}
    \caption{Best-fit EFT predictions against the simulated power spectra (left panel),
    and the residuals between the model and the data (right panel). 
    The best-fit model was obtained in the 3-spectra analysis for the most massive halos with $(\kmax^\lya,\kmax^\qso,\kmax^\cross)=(2,0.8,1)\hMpc$.
    The constant shot-noise contribution is subtracted from the $P^\qso$ data.}
    \label{fig:bestfit}
\end{figure*}

The best-fit predictions for our baseline 3-spectra model across four angular bins are compared to the data in Fig.~\ref{fig:bestfit}.
The EFT model predicts the $P^\lya$, $P^\qso$ and $P^\cross$ data at $(\kmax^\lya,\kmax^\qso,\kmax^\cross)=(2,0.8,1)\hMpc$ with 
the nominal 
$0.8\%$, $3.2\%$ and $0.4\%$ accuracy, respectively.
The residuals grow 
at lower $k$
due to the significant 
cosmic variance
of the Sherwood simulation. 
These results represent a significant improvement over the previous analysis~\cite{Givans:2022qgb}, which described the Ly$\alpha$ -- halo cross-power spectrum data with a $10\%$ error up to scales of $\kmax^\cross = 1\hMpc$.

As an extension of our analysis, 
we add $P_{\rm shot}^\cross$ to the fit
at $\kmax^\cross=1~\hMpc$. We do not 
detect this parameter in the data. 
Our $68\%$ constraint is given by
\[
\frac{P^\cross_{\rm shot}}{[\Mpch]^3}= -1.15\pm 1.83\,.
\]
Additionally, we checked in post-processing 
that adding $P^\cross_{\rm shot}$
to the best-fit models from other 
analyses does not improve the fit
and does not increase the 
accuracy of the EFT model at higher
$\kmax$. These analyses validate our
baseline choice $P^\cross_{\rm shot}=0$.

\subsubsection{$P^\lya+P^\qso+P^\cross$ analysis: off-diagonal covariance}

We now present the results of the 3-spectra analysis using the full off-diagonal covariance \eqref{cov}, \eqref{covL}.

\begin{figure*}
    \centering
    \includegraphics[width=0.99\textwidth]{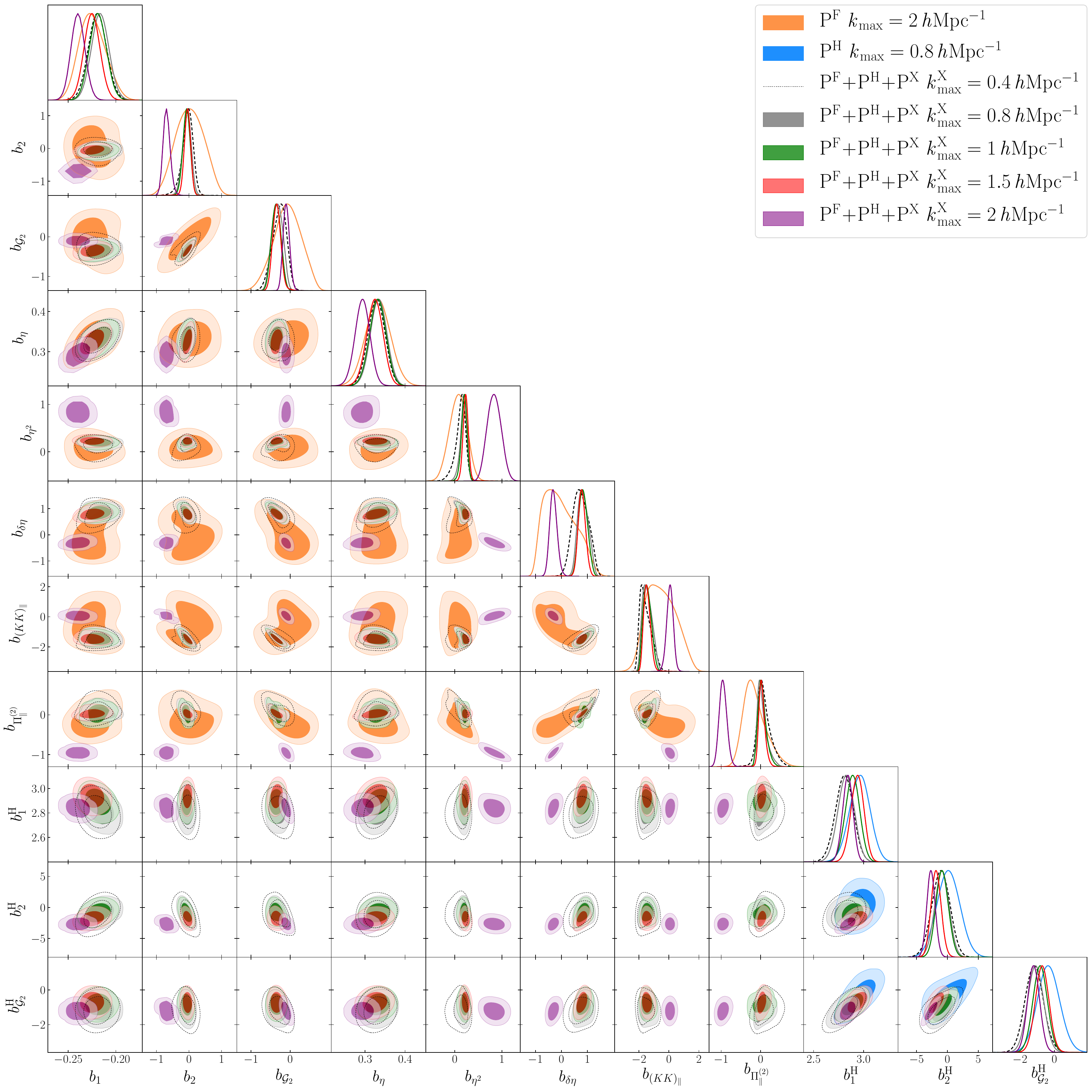}\\
    \caption{Marginalized posteriors on nuisance parameters of the EFT model for the Ly$\alpha$ forest auto-power spectrum $P^\lya$ (orange), the massive halo auto-power spectrum $P^\qso$ (blue), and their combination with the Ly$\alpha$ -- halo cross-power spectrum $P^\cross$ at $z=2.8$. The combined analysis results are shown for four different values of $k^\cross_{\rm max}$: $0.4$, $0.8$, $1$, $1.5$, and $2\,\hMpc$ (dashed black, gray, green, red, purple, respectively). All results are obtained with the off-diagonal covariance.}
    \label{fig:all_off}
\end{figure*}

Fig.~\ref{fig:all_off} shows the 1D and 2D marginalized posterior distributions for the bias parameters derived from various data combinations.
The results are shown for five different $\kmax^\cross$ values: $0.4$, $0.8$, $1$, $1.5$, and $2\hMpc$.
Similarly to the analysis with the diagonal covariance, we observe that the constraints obtained from the $P^\lya$ and $P^\qsoo$ individually are fully consistent with those from the combined 3-spectra analysis up to $\kmax^\cross=1.5\hMpc$.
However, the constraints on the parameters are significantly tighter and posteriors shift more rapidly as $\kmax^\cross$ increases compared to analysis with the diagonal covariance, cf. with Fig.~\ref{fig:all}.
This indicates that the off-diagonal terms in the covariance introduce additional constraints, leading to a worse fit.

As a frequency confirmation of our results, we observe a worse fit compared to the analysis using a diagonal covariance. 
Specifically, for $\kmax^\cross = 1\hMpc$, the minimum $\chi^2$ value is $434$ for $355$ data points, with a similar fit quality observed for $\kmax^\cross = 0.4$ and $0.8\hMpc$. 
This degradation in the fit quality can be explained by the additional correlations between data points introduced by the off-diagonal terms in the covariance.
As $\kmax^\cross$ increases, the fit quality gradually deteriorates: for $\kmax^\cross = 1.5$, the $\chi^2$ value rises to $466$ for $371$ data points; for $\kmax^\cross = 2$ it further increases to $497$ for $387$ data points.
Given that the fit quality for $\kmax^\cross \leq1\hMpc$ is comparable and to ensure a direct comparison with the results in
Sec.~\ref{sec:comb_HM}, we choose $\kmax^\cross = 1\hMpc$ as a baseline for the 3-spectra analysis with off-diagonal covariance.

Specifically, for $\kmax^\cross = 1\hMpc$, the minimum $\chi^2$ value is $434$ for $355$ data points, with a similar fit quality observed for $\kmax^\cross = 0.4$ and $0.8\hMpc$. 
With 28 free parameters, these results indicate similarly poor fits.
As $\kmax^\cross$ increases, the fit quality gradually deteriorates: for $\kmax^\cross = 1.5$, the $\chi^2$ value rises to $466$ for $371$ data points; for $\kmax^\cross = 2$ it further increases to $497$ for $387$ data points.
Given that the fit quality for $\kmax^\cross \leq1\hMpc$ is comparable and to ensure a direct comparison with the results in
Sec.~\ref{sec:comb_HM}, we choose $\kmax^\cross = 1\hMpc$ as a baseline for the 3-spectra analysis with off-diagonal covariance.

Tab.~\ref{tab:all} compares the constraints on EFT parameters obtained when using the diagonal covariance (fifth column) and the off-diagonal covariance (forth column).
For the $\lyaa$ forest linear and quadratic bias parameters, the improvement in constraints ranges from $10\%$ to $50\%$ relative to the analysis with diagonal covariance.
A similar level of enhancement is found for the halo parameters, with the most significant impact on $\tilde c^\qso$, whose error is reduced by a factor of 2.
Notably, the next-to-leading order counterterm for the Ly$\alpha$ -- halo cross-power spectrum $\tilde c^\cross$ is nonzero at the $2.7\sigma$ significance level that motivates its inclusion in the analysis.
Overall, the constraints from the baseline analyses with diagonal and off-diagonal covariances are fully consistent with each other. This validates our analysis procedure, including the adopted data cuts.

\begin{figure}[!t]
    \centering
    \includegraphics[width=0.5\textwidth]{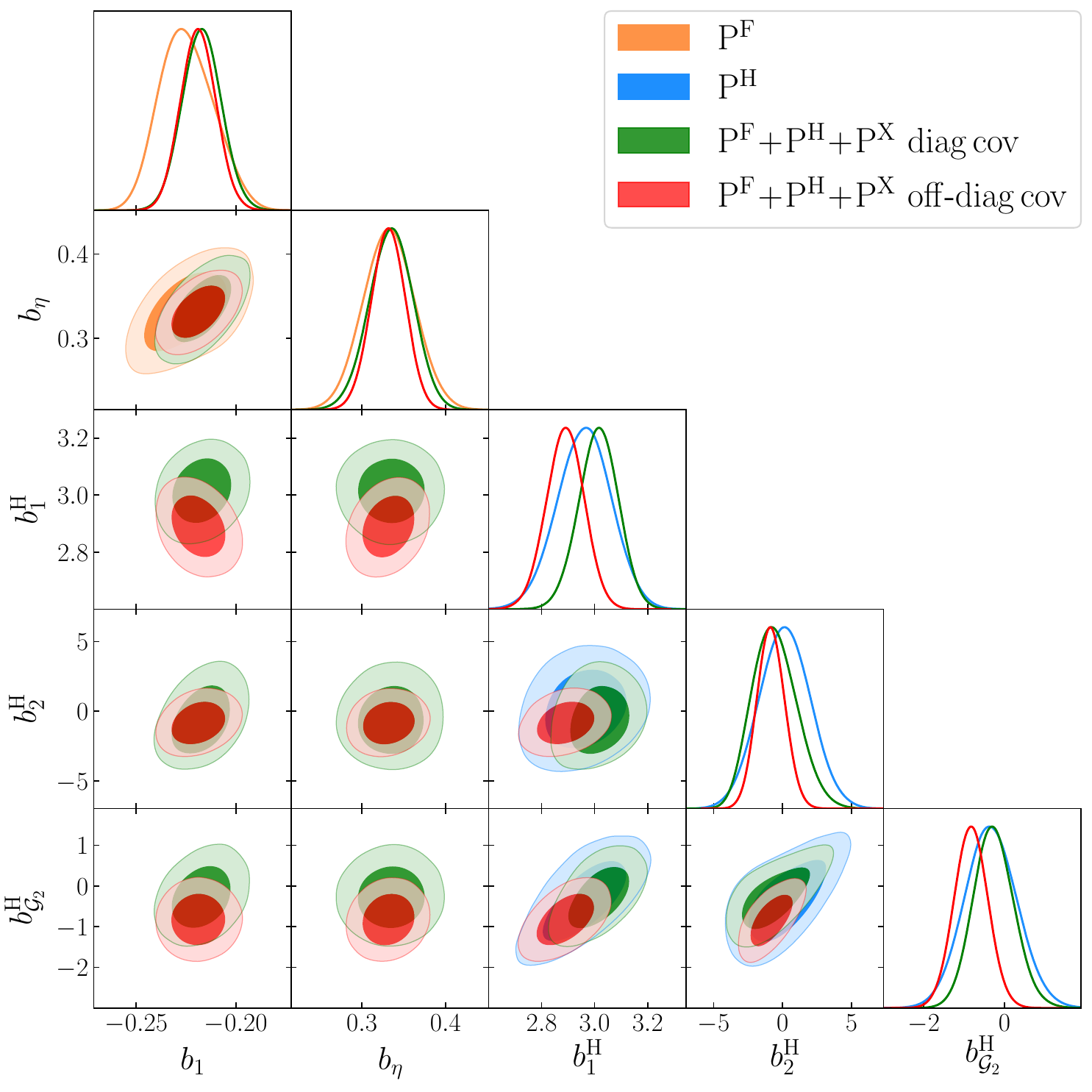}\\
    \caption{Marginalized posteriors obtained from the Ly$\alpha$ forest auto-power spectrum with $\kmax=2\hMpc$ (orange), the massive halo auto-power spectrum with $\kmax=0.8\hMpc$ (blue), and their combination with the Ly$\alpha$ -- halo cross-power spectrum with $\kmax^\cross=1\hMpc$, using the diagonal
    covariance (green) and the off-diagonal covariance (red). 
    }
    \label{fig:HM_diffcov}
\end{figure}

Fig.~\ref{fig:HM_diffcov} compares the posteriors from the 3-spectra analyses with the diagonal and off-diagonal covariance. 
We observe that our parameter constraints are fully consistent between the two cases, with the analysis using off-diagonal covariance providing significantly tighter constraints.
The analysis with the full off-diagonal covariance yields robust measurements, but it significantly deteriorates the fit quality if one compares the $\chi^2$ statistics
with the number of Fourier bins (minus the number of fitting parameters). 
This comparison, however, may 
be 
misleading as the cross-covariance explicitly 
correlates these bins.
In addition, the naive counting 
of EFT parameters as degrees of 
freedom is misleading too as often these parameters are highly degenerate (e.g. $b_{\mathcal{G}_2}$ and $b_{\Gamma_3}$), so that an 
addition of extra parameters does not necessarily increase the flexibility
of the fit.

\subsection{All halos}
\label{sec:res2}

\subsubsection{$P^\qsoo$ analysis}
\label{sec:res2_1}

In this section, we present the results obtained from the complete halo catalog.
We start with the analysis of the auto-power spectrum of light halos.

\begin{figure}[!t]
    \centering
    \includegraphics[width=0.5\textwidth]{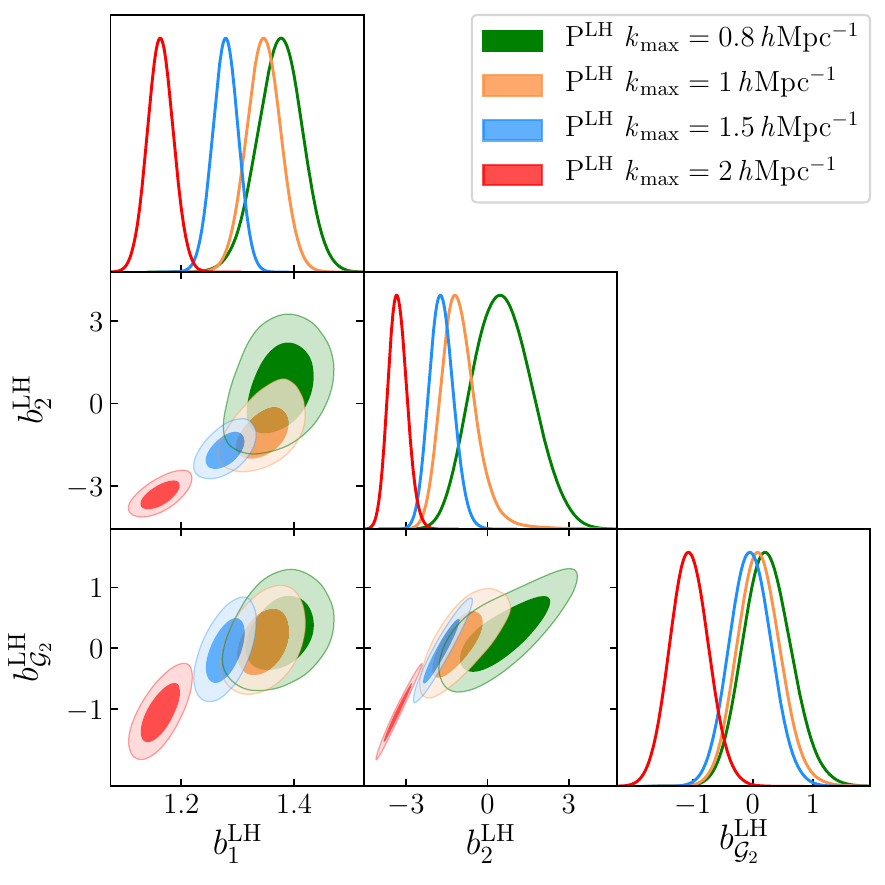}\\
    \caption{Marginalized posteriors obtained from the $P^\qsoo$ for different values of $k_{\rm max}$: $0.8$, $1$, $1.5$ and $2\,h\Mpc^{-1}$ (green, orange, blue, red, respectively).}
    \label{fig:haloAH}
\end{figure}

Fig.~\ref{fig:haloAH} shows the posterior distributions for the bias parameters.
The constraints on the bias parameters appear to be tighter compared to those from the $P^\qso$ analysis, cf. with Fig.~\ref{fig:halo}.
This improvement can be attributed to the higher number density of the full halo catalog. 
Consequently, as $\kmax$ increases, the contours shifts more rapidly, implying that the perturbative approach for light halos breaks down at smaller scales.
For $\kmax = 1\hMpc$, the best-fit $\chi^2$ statistics shows a good agreement with the data.
However, for smaller data cuts, the fit quality deteriorates substantially, leading to shifts in the posteriors.
For instance, for $\kmax=2\hMpc$, the parameter $b_2^\qsoo=-3.3\pm0.4$ shows a significant tension with the peak-background split result $\tilde b_2^\qsoo=-0.5$~\cite{Lazeyras:2015lgp} and 
the measurements of~\cite{Ivanov:2024dgv}, suggesting a biased fit.~\footnote{See footnote~\ref{fn:split}.}

We evaluate the magnitude of the one-loop correction.
Our analysis shows that one-loop contribution remains below $20\%$ of the linear theory prediction up to scales of $\kmax=0.8\hMpc$.
Thus, we select a conservative scale cut $\kmax=0.8\hMpc$ as a baseline for the all-halo power spectrum analysis.
This choice also justifies the scale cut used in the $P^\qso$ analysis in Sec.~\ref{sec:res1}, as the full halo catalog, with its greater statistical power, offers a more stringent test of the theoretical model.
Additionally, using the same scale cut $\kmax$ for both the light and massive halos facilitates a direct comparison of the results.

\subsubsection{$P^\lya+P^\qsoo+P^\cross$ analysis: diagonal covariance}
\label{sec:res2_2}

We now present the parameter constraints from the combined $P^\lya+P^\qsoo+P^\cross$ analysis. 
Consistent with the analysis of the most massive halos, we fix $\kmax^\lya = 2\hMpc$ and $\kmax^\qso = 0.8\hMpc$, and vary $\kmax^\cross$ independently.

\begin{figure*}
    \centering
    \includegraphics[width=0.99\textwidth]{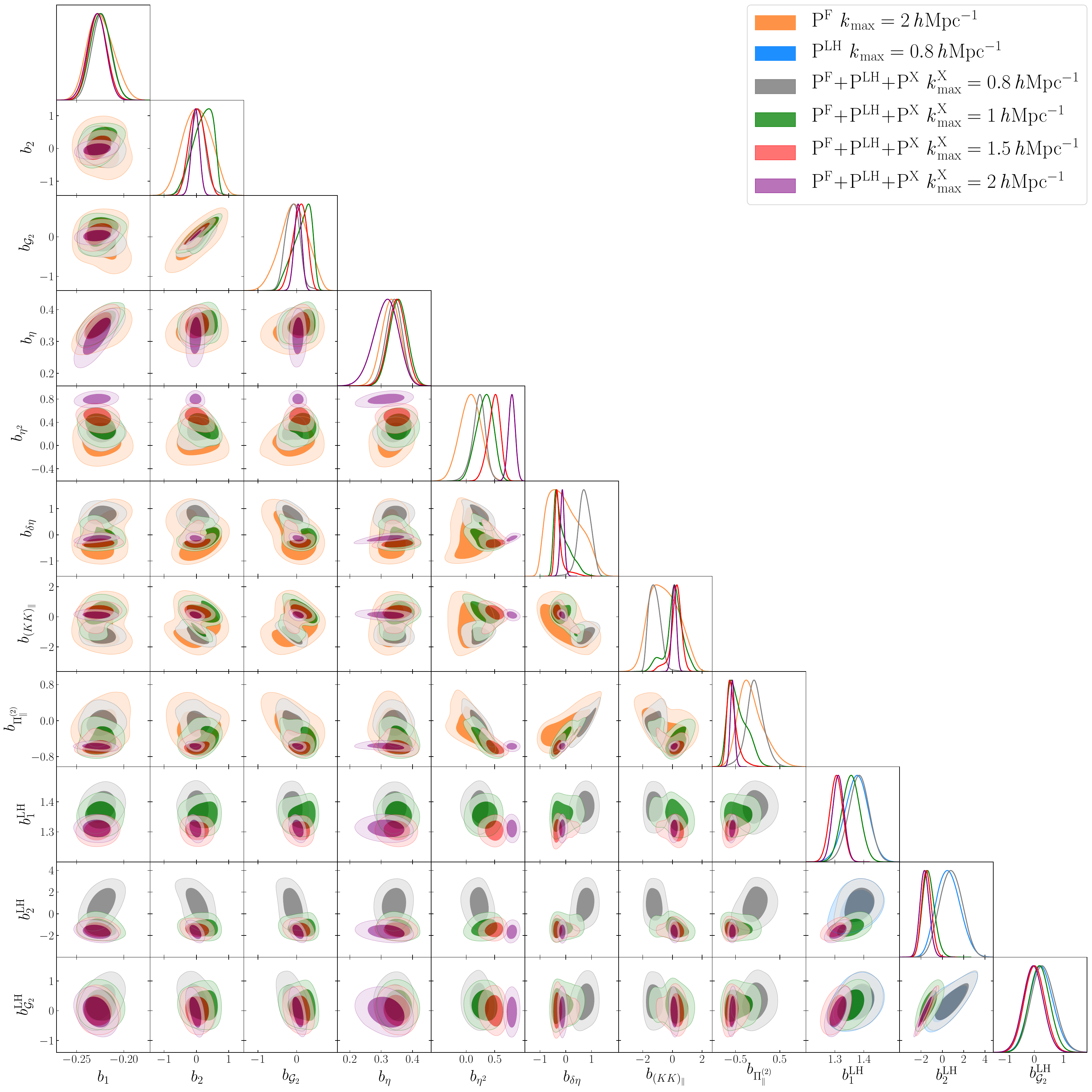}\\
    \caption{Marginalized posteriors on nuisance parameters of the EFT model for the Ly$\alpha$ forest auto-power spectrum $P^\lya$ (orange), the all-halo auto-power spectrum $P^\qsoo$ (blue), and their combination with the Ly$\alpha$ -- all-halo cross-power spectrum $P^\cross$ at $z=2.8$. The combined analysis results are shown for four different values of $k^\cross_{\rm max}$: $0.8$, $1$, $1.5$, and $2\,\hMpc$ (gray, green, red, purple, respectively). The results are obtained with the diagonal covariance.}
    \label{fig:all_AH}
\end{figure*}

Fig.~\ref{fig:all_AH} shows the 1D and 2D marginalized posterior distributions for the bias parameters derived from various data combinations.
The constraints derived from $P^\lya$ and $P^\qsoo$ individually are entirely consistent with those from the 3-spectra analysis up to $\kmax^\cross=1\hMpc$.
For $\kmax^\cross>1\hMpc$, the contours for the bias parameters $b_{\eta^2}$ and $b_{\Pi^{[2]}_\parallel}$ shift progressively.
This suggests that the fit is biased at $\kmax^\cross=1.5\hMpc$.

To confirm our scale cuts, we perform a $\chi^2$ test for the combined 3-spectra analysis.
For $\kmax^\cross=1\hMpc$, the best-fit $\chi^2$ statistic is $376$ for the $355$ data points, indicating a good fit for 28 free parameters.
For $\kmax^\cross>1\hMpc$, the fit quality deteriorates: for $\kmax^\cross = 1.5$, the $\chi^2$ values increase to $409$ for $371$ data points; for $\kmax^\cross = 2$ it rises further to $459$ for $387$ data points.
These findings support the choice of $\kmax^\cross=1\hMpc$ as our baseline scale cut.

The 1D marginalized parameter constraints for the baseline configurations are given in Tab.~\ref{tab:all_AH}. 
\begin{table*}[!htb]
\renewcommand{\arraystretch}{1.2}
%\normalsize
\small
\centering
\begin{tabular}{|l||c|c|c c|}
 \hline
\multirow{3}{*}{\diagbox{ {\small Param.}}{\small Data}}
& \multirow{3}{*}{$P^\lya$}
& \multirow{3}{*}{$P^\qsoo$} 
& \multicolumn{2}{c|}{\multirow{2}{*}{$P^\lya+P^\qsoo+P^\cross$}}  
%& $P^\lya+P^\qso+P^\cross$ 
\\ 
& & & & 
\\
%\cline{4-5}
& & & \text{diag cov} & \text{off-diag cov}
\\
\hline\hline
%lya
$b_{1 }$ 
& $-0.2245_{-0.0154}^{+0.0126}$ %
& --
& $-0.2244_{-0.0108}^{+0.0105}$ %% 
& $-0.2254_{-0.0078}^{+0.0076}$ %%
\\
$b_\eta$ 
& $0.332_{-0.031}^{-0.031}$ %
& --
& $0.355_{-0.029}^{+0.030}$ %%  
& $0.346_{-0.014}^{+0.014}$ %%   
\\
$b_{2 }$ 
& $0.03_{-0.46}^{+0.44}$ %
& --
& $0.21_{-0.23}^{+0.39}$ %%
& $0.61_{-0.05}^{+0.05}$ %% 
\\
$b_{\mathcal{G}_2 }$ 
& $-0.07_{-0.34}^{+0.38}$ %
& --
& $0.14_{-0.17}^{+0.30}$ %%
& $0.46_{-0.04}^{+0.04}$ %% 
\\
$b_{\eta^2}$  
& $0.072_{-0.180}^{+0.194}$ %
& --
& $0.319_{-0.139}^{+0.169}$ %% 
& $0.304_{-0.072}^{+0.071}$ %%  
\\
$b_{\delta\eta}$  
& $-0.03_{-0.83}^{+0.42}$ %
& --
& $-0.12_{-0.41}^{+0.15}$ %% 
& $0.07_{-0.11}^{+0.09}$ %%  
\\
$b_{(KK)_\parallel}$ 
& $-0.51_{-1.18}^{+0.86}$ %
& --
& $0.10_{-0.40}^{+0.69}$ %% 
& $0.04_{-0.14}^{+0.10}$ %%
\\
$b_{\Pi^{[2]}_\parallel}$  
& $-0.142_{-0.350}^{+0.212}$ %
& --
& $-0.442_{-0.290}^{+0.145}$ %%
& $-0.326_{-0.085}^{+0.070}$ %% 
\\
\hline
$b_{\Gamma_3}$ 
& $-0.49_{-0.13}^{+0.13}$ %
& --
& $-1.42_{-0.11}^{+0.11}$ %%
& $-1.82_{-0.10}^{+0.10}$ %%
\\
$10^2 c_0/[\Mpch]^2$ 
& $-2.32_{-1.06}^{+1.06}$  %
& --
& $-1.76_{-0.26}^{+0.26}$ %% 
& $-1.85_{-0.25}^{+0.25}$ %%  
\\
$10^2 c_2/[\Mpch]^2$ 
& $4.26_{-1.54}^{+1.54}$ %
& --
& $3.09_{-1.01}^{+1.01}$ %% 
& $4.37_{-0.94}^{+0.94}$ %%  
\\
$10^2 c_4/[\Mpch]^2$ 
& $-5.38_{-1.01}^{+1.01}$ %
& -- 
& $-4.87_{-0.84}^{+0.84}$ %% 
& $-7.55_{-0.84}^{+0.84}$ %%  
\\
$b_{\Pi^{[3]}_\parallel}$ 
& $0.771_{-0.089}^{+0.089}$ %
& --
& $1.322_{-0.084}^{+0.084}$ %%
& $1.272_{-0.071}^{+0.071}$ %% 
\\
$b_{\delta\Pi^{[2]}_\parallel}$ 
& $-0.05_{-0.19}^{+0.19}$ %
& -- 
& $-0.75_{-0.18}^{+0.18}$ %% 
& $-1.33_{-0.17}^{+0.17}$ %%  
\\
$b_{(K\Pi^{[2]})_\parallel}$ 
& $-1.64_{-0.25}^{+0.25}$ %
& --
& $-2.66_{-0.20}^{+0.20}$ %% 
& $-2.97_{-0.20}^{+0.20}$ %%  
\\
$b_{\eta\Pi^{[2]}_\parallel}$ 
& $-0.24_{-0.43}^{+0.43}$ % 
& --
& $0.80_{-0.42}^{+0.42}$ %%
& $-1.83_{-0.42}^{+0.42}$ %% 
\\
\hline\hline
%qso
$b_{1 }^\qsoo$ 
& -- 
& $1.375_{-0.039}^{+0.042}$ % 
& $1.356_{-0.032}^{+0.031}$ %%
& $1.324_{-0.028}^{+0.028}$ %% 
\\
$b_{2 }^\qsoo$ 
& -- 
& $0.58_{-1.20}^{+1.03}$ %
& $-1.25_{-0.68}^{+0.52}$ %%
& $-1.09_{-0.50}^{+0.45}$ %% 
\\
$b_{\mathcal{G}_2 }^\qsoo$ 
& --
& $0.25_{-0.44}^{+0.39}$ %  
& $0.19_{-0.38}^{+0.36}$ %% 
& $0.04_{-0.31}^{+0.30}$ %%  
\\
\hline
$b_{\Gamma_3}^\qsoo$ 
& --
& $0.14_{-0.13}^{+0.13}$ %
& $-0.34_{-0.07}^{+0.07}$ %%
& $-0.71_{-0.06}^{+0.06}$ %% 
\\
$c_0^\qsoo/[\Mpch]^2$ 
& --
& $0.024_{-0.087}^{+0.087}$ %
& $-0.223_{-0.035}^{+0.035}$ %% 
& $-0.058_{-0.025}^{+0.025}$ %%   
\\
$c_2^\qsoo/[\Mpch]^2$ 
& --
& $1.404_{-0.159}^{+0.159}$ % 
& $0.183_{-0.088}^{+0.088}$ %% 
& $0.255_{-0.049}^{+0.049}$ %%  
\\
$c_4^\qsoo/[\Mpch]^2$ 
& --
& $2.153_{-0.419}^{+0.419}$ % 
& $1.60_{-0.22}^{+0.22}$ %% 
& $2.05_{-0.11}^{+0.11}$ %%
\\
$\tilde{c}^\qsoo/[\Mpch]^4$ 
& --
& $-0.81_{-0.77}^{+0.77}$ % 
& $0.65_{-0.40}^{+0.40}$ %% 
& $-0.20_{-0.09}^{+0.09}$ %%  
\\
$P_\sh^\qsoo$ 
& --
& $0.001_{-1.0}^{+1.0}$ %
& $0.02_{-0.86}^{+0.86}$ %%
& $-0.11_{-0.32}^{+0.32}$ %%
\\
$a_0^\qsoo$ 
& --
& $0.0_{-1.0}^{+1.0}$ %
& $0.0_{-1.0}^{+1.0}$ %%
& $0.02_{-1.0}^{+1.0}$ %%
\\
$a_2^\qsoo$ 
& --
& $0.0_{-1.0}^{+1.0}$ %
& $-0.01_{-1.0}^{+1.0}$ %%
& $0.01_{-1.0}^{+1.0}$ %%
\\
\hline\hline
%cross
$\tilde{c}^\cross/[\Mpch]^4$ 
& --
& --
& $-0.086_{-0.104}^{+0.104}$ %%
& $-0.253_{-0.046}^{+0.046}$ %%
\\
 \hline
 \end{tabular} 
 \caption{One-dimensional marginalized constraints on 
 nuisance parameters of the one-loop EFT model from the Ly$\alpha$ forest auto-power spectrum with $\kmax^\lya=2\,\hMpc$ (second column), the all-halo auto-power spectrum with $\kmax^\qso=0.8\,\hMpc$ (third column) and their combination with the Ly$\alpha$ -- all-halo cross-power spectrum with $\kmax^\cross=1\,\hMpc$, using diagonal covariance (fourth column) and off-diagonal covariance (fifth column), at $z=2.8$. Parameter constraints related to each respective spectrum are grouped together. The parameters in the upper section were directly sampled in our MCMC chains, while the parameters in the lower section were analytically marginalized in the likelihood, with their posteriors recovered from the chains~\textit{a posteriori}.
 \label{tab:all_AH}}
\end{table*}
The inclusion of $P^\cross$ significantly enhances the constraints obtained from the $P^\lya$ and $P^\qsoo$ individually.
For the Lyman-$\alpha$ forest bias parameters, the errors are reduced by factors ranging from $1.2$ to $2.2$ when compared to the $P^\lya$-only analysis.
Unlike the analysis of the most massive halos, the uncertainties on the cubic bias parameters decrease compared to the $P^\lya$-only case, though the overall improvement is modest.
Additionally, the uncertainty on $c_0$ is reduced by more than a factor of $4$, while the gains for $c_2$, $c_4$ are more moderate. 
The constraints on the halo bias parameters exhibit significant improvement, particularly for $b_2^\qsoo$ and $b_{\Gamma_3}^\qsoo$, whose errors are reduced by nearly a factor of 2.
The improvement for the halo counterterms ranges from $1.8$ to $2.5$ times.
We conclude that the $P^\lya$, $P^\qsoo$, and $P^\cross$ measurements are highly complementary, though the information gain is somewhat reduced compared to the analysis of the most massive halos.
This reduction can be attributed to the greater statistical power of the light halo power spectrum, which diminishes the impact of the cross-correlation in the 3-spectra analysis.

\begin{figure*}
    \centering
    \includegraphics[width=0.41\textwidth]{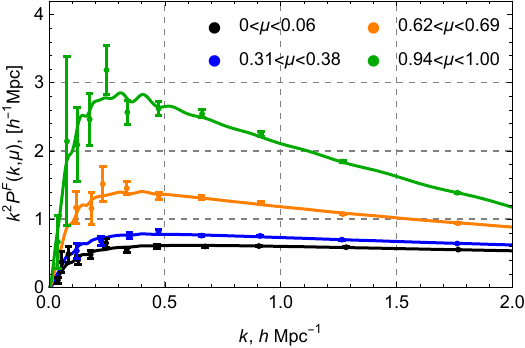}
    \includegraphics[width=0.41\textwidth]{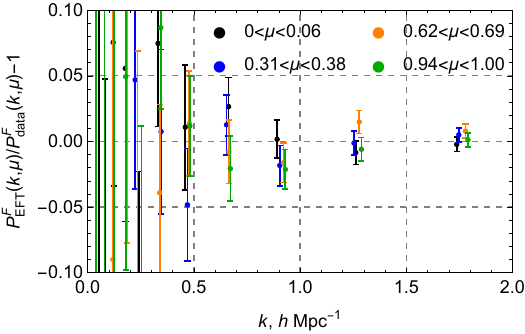}\\
    \includegraphics[width=0.41\textwidth]{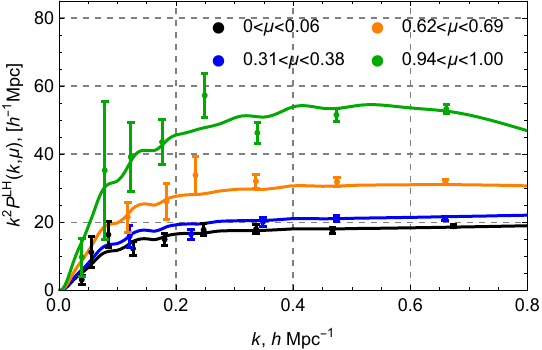}
    \includegraphics[width=0.41\textwidth]{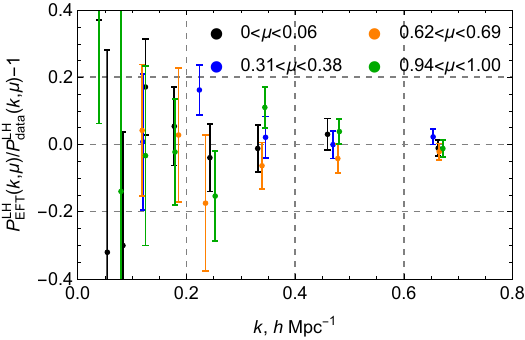}\\
    \includegraphics[width=0.41\textwidth]{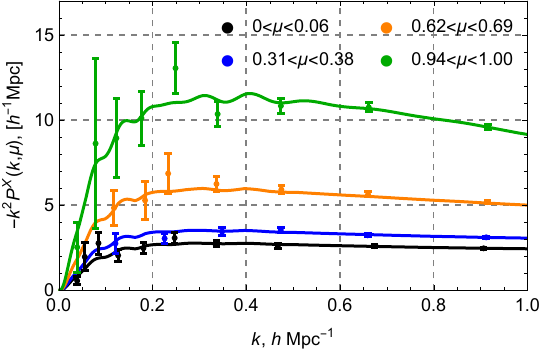}
    \includegraphics[width=0.41\textwidth]{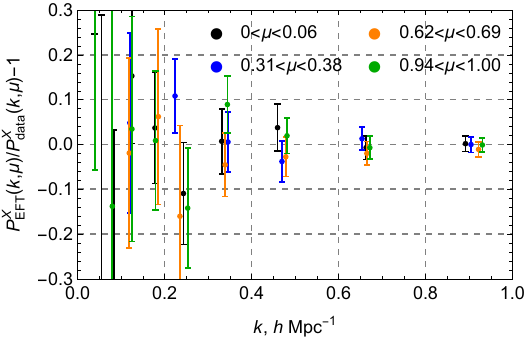}
    \caption{Best-fit EFT predictions against the simulated power spectra (left panel),
    and the residuals between the model and the data (right panel). The best-fit model was obtained in the 3-spectra analysis for light halos with $(\kmax^\lya,\kmax^\qso,\kmax^\cross)=(2,0.8,1)\hMpc$. The constant shot-noise contribution is subtracted from the $P^\qsoo$ data.}
    \label{fig:bestfit_AH}
\end{figure*}

The best-fit predictions for our baseline 3-spectrum model across four angular bins are shown in Fig.~\ref{fig:bestfit_AH}.
The EFT model describes the $P^\lya$, $P^\qsoo$ and $P^\cross$ data at $(\kmax^\lya,\kmax^\qso,\kmax^\cross)=(2,0.8,1)\hMpc$ with $0.8\%$, $2.2\%$ and $1.1\%$ accuracy, respectively.
We observe a moderate improvement in the modeling of the halo auto-power spectrum relative to the analysis in Sec.~\ref{sec:comb_HM}.
This improvement can be attributed to the greater statistical power of the full halo catalog, which provides more precise measurements. 
Our findings demonstrate a fivefold improvement in modeling the Ly$\alpha$ -- halo cross-power spectrum compared to the earlier analysis in ref.~\cite{Givans:2022qgb}, which only achieved $5\%$ accuracy up to $\kmax^\cross = 1\hMpc$ for light halos.

Finally, as an extension of our analysis, 
we add $P_{\rm shot}^\cross$ to the fit
at $\kmax^\cross=1~\hMpc$. We do not 
detect this parameter in the data. 
Our $68\%$ constraint reads
\[
\frac{P^\cross_{\rm shot}}{[\Mpch]^3}= 0.13\pm 0.24\,.
\]
We have additionally checked in post-processing 
that adding $P^\cross_{\rm shot}$
to the best-fit models from other 
analyses does not improve the fit
and does not increase the 
accuracy of the EFT model at higher
$\kmax$. These analyses validate our
baseline choice $P^\cross_{\rm shot}=0$.

\subsubsection{$P^\lya+P^\qsoo+P^\cross$ analysis: off-diagonal covariance}

We now proceed to the 3-spectra analysis with the full off-diagonal covariance.  

\begin{figure*}
    \centering
    \includegraphics[width=0.99\textwidth]{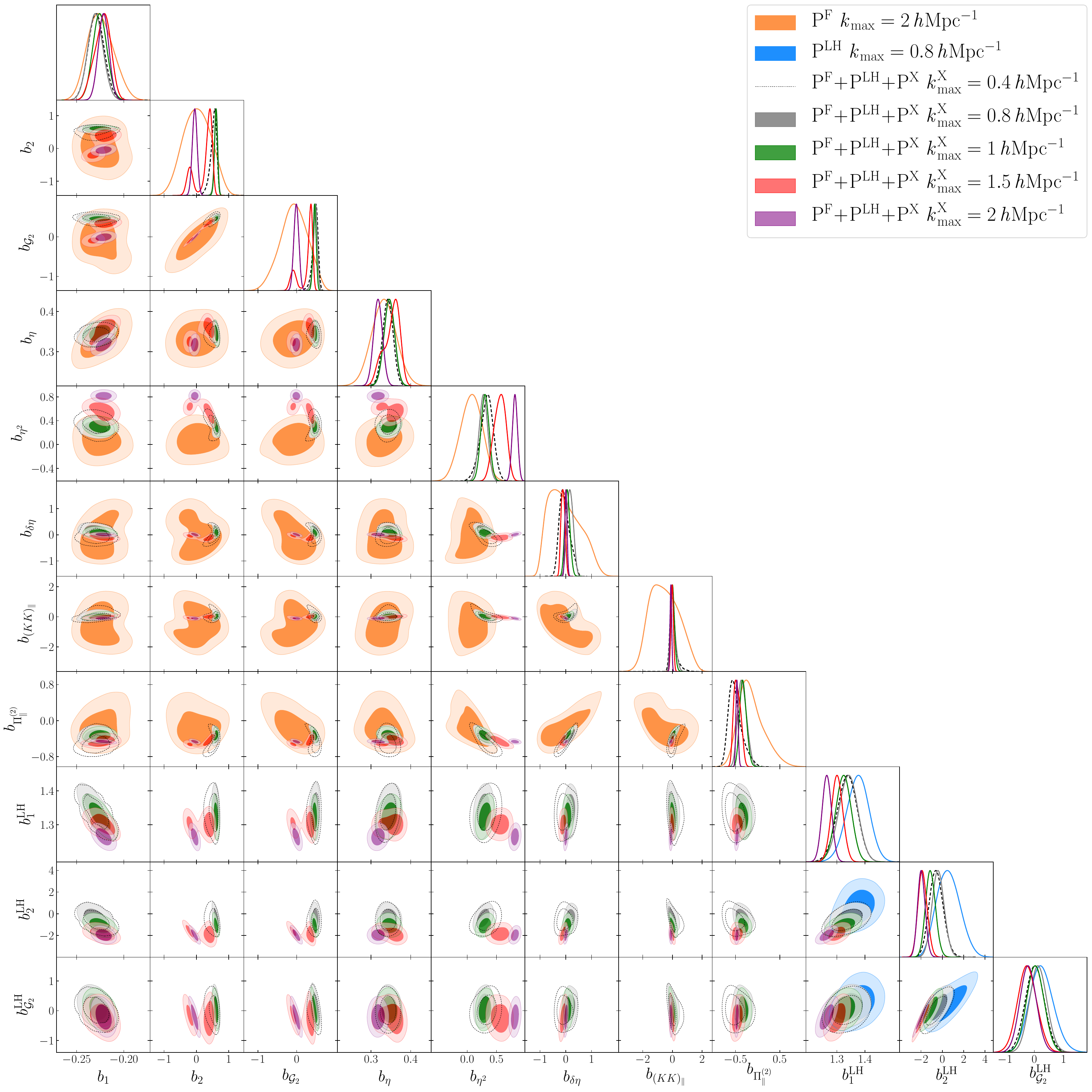}\\
    \caption{Marginalized posteriors on nuisance parameters of the EFT model for the Ly$\alpha$ forest auto-power spectrum $P^\lya$ (orange), the all-halo auto-power spectrum $P^\qsoo$ (blue), and their combination with the Ly$\alpha$ -- all-halo cross-power spectrum $P^\cross$ at $z=2.8$. The combined analysis results are shown for four different values of $k^\cross_{\rm max}$: $0.4$, $0.8$, $1$, $1.5$, and $2\,\hMpc$ (dashed black, gray, green, red, purple, respectively). The results are obtained with the off-diagonal covariance.}
    \label{fig:all_AH_off}
\end{figure*}

Fig.~\ref{fig:all_AH_off} shows the 1D and 2D marginalized posterior distributions for the bias parameters derived from various data combinations.
Up to $\kmax^\cross=1\hMpc$, the constraints from the combined 3-spectra analysis are perfectly consistent with those obtained from 
the $P^\lya$ and $P^\qsoo$ individually.  
Similar to the analysis with the massive halos, the posteriors are significantly tightened compared to the results with diagonal covariance.
For $\kmax^\cross>1\hMpc$, the posteriors shift progressively, indicating a biased fit at these scales.

Similar to the analysis with the massive halos, the fit quality is relatively poor at $\kmax^\cross\leq1\hMpc$. 
Specifically, for $\kmax^\cross=1\hMpc$ the minimum $\chi^2$ value is $485$ across the $355$ data points.
As $\kmax^\cross$ increases, the fit quality gradually deteriorates: for $\kmax^\cross = 1.5$, the $\chi^2$ value increases to $520$ for $371$ data points, and for $\kmax^\cross=2\hMpc$ it rises sharply to $603$ for $387$ data points.
Given that the fit quality for
$\kmax^\cross \leq1\hMpc$
is comparable and to be aligned with the analysis in Sec.~\ref{sec:res2_2}, we select $\kmax^\cross=1\hMpc$ as a baseline for the 3-spectra analysis with the off-diagonal covariance.

Tab.~\ref{tab:all_AH} compares the constraints on EFT parameters obtained when using diagonal covariance and off-diagonal covariance.
The uncertainties on the $\lyaa$ forest linear and quadratic bias parameters decrease substantially -- by factors ranging from $1.4$ to $6.5$ -- when off-diagonal covariance is used.
For the halo bias parameters, the improvement is more modest.
The improvement for the halo counterterms ranges from $1.4$ to $4.5$ times.
Importantly, the next-to-leading order counterterm for the Ly$\alpha$ -- halo cross-power spectrum $\tilde c^\cross$ is detected with high significance that motivates its inclusion in the analysis.
Overall, the improvement in constraints with the off-diagonal covariance is somewhat larger for light halos as compared to the massive halos, cf. Tab.~\ref{tab:all}.

\begin{figure}[!t]
    \centering
    \includegraphics[width=0.5\textwidth]{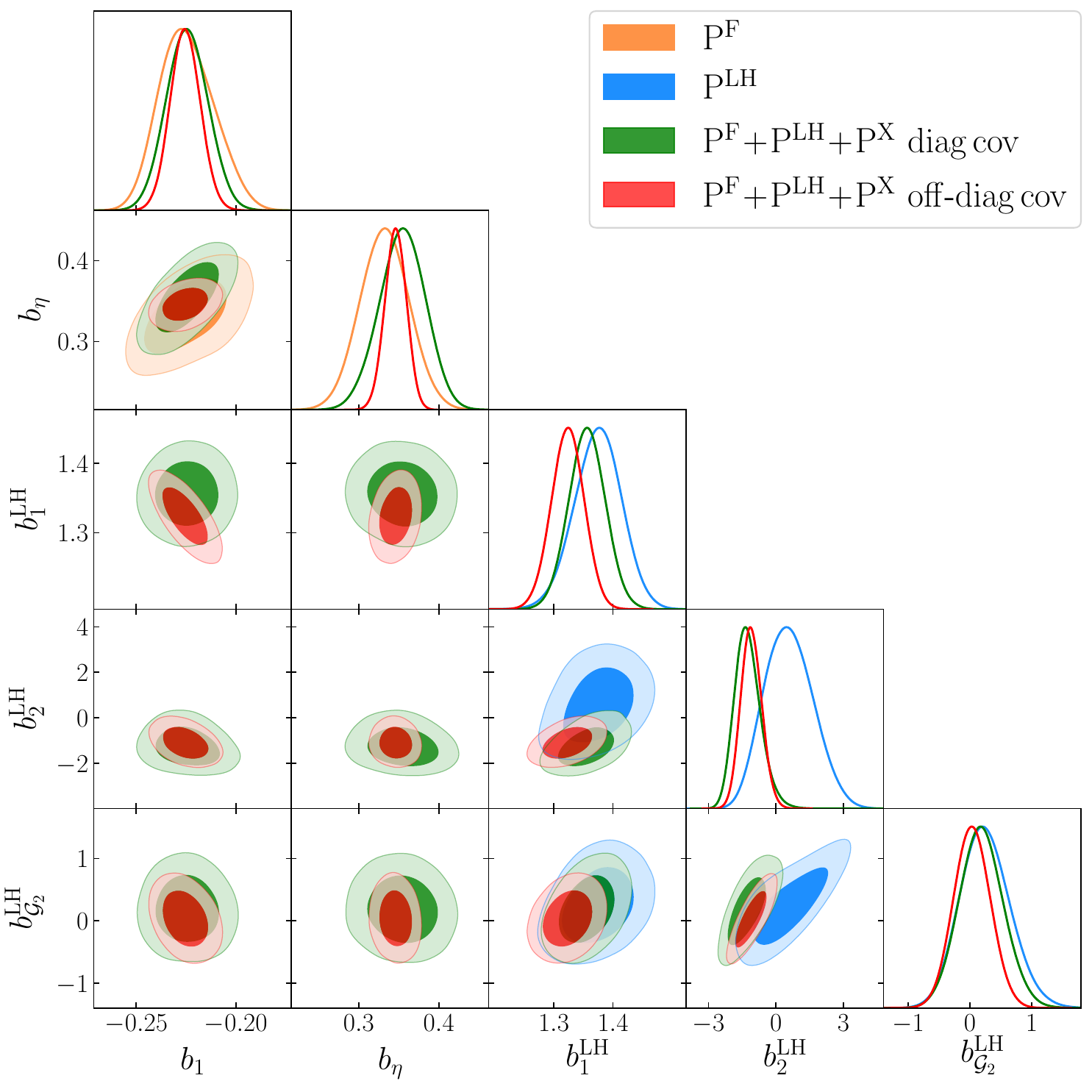}\\
    \caption{Marginalized posteriors obtained from the Ly$\alpha$ forest auto-power spectrum with $\kmax=2\hMpc$ (orange), the all-halo auto-power spectrum with $\kmax=0.8\hMpc$ (blue), and their combination with the Ly$\alpha$ -- all-halo cross-power spectrum with $\kmax^\cross=1\hMpc$, using diagonal
    covariance (green) and off-diagonal covariance (red). 
    }
    \label{fig:diffcov}
\end{figure}

Fig.~\ref{fig:diffcov} compares the posteriors from the 3-spectra analyses with diagonal and off-diagonal covariances. 
The parameter constraints are consistent between the two analyses; however, using off-diagonal covariance results in significantly tighter constraints.
Unlike the case of massive halos, the analysis with the off-diagonal covariance considerably improves the constraints on 
the linear bias parameters.
For instance, the error on $b_\eta$ is reduced by more than a factor of $2$ compared to the analysis with the diagonal covariance.

\section{Summary and Conclusions}
\label{sec:conc}

We have presented the one-loop EFT model for the 
cross spectrum of the $\La$ forest and a generic biased tracer of matter.
Our work is primarily aimed at the analytic description 
of the $\La$ forest -- galaxy and $\La$ forest -- quasar cross correlations.
For practical comparisons, 
we use massive halos 
from simulations 
as a proxy for quasars,
and light halos as a proxy 
for abundant 
high-redshift galaxies
such as Lyman-$\alpha$
emitters. 
We have found an excellent agreement between our theoretical model
and the $\La$-halo data from the Sherwood hydrodynamical 
simulation data on quasi-linear scales. Specifically, the 1-loop EFT model
provides a percent-level accuracy in fitting these data at $\kmax= 1~\hMpc$.
Our main results are displayed
in figs.~\ref{fig:HM_diffcov}
and~\ref{fig:diffcov} for massive and light halos, respectively.  
Our model can be readily applied to the cross-correlation data 
from eBOSS and DESI surveys.

We have also found that the 
results depend noticeably
on the covariance used. In particular, the inclusion of the
analytical 
Gaussian 
off-diagonal covariance
between the halo, $\La$ flux, 
and the cross spectra 
leads to significantly 
improved constraints
on the EFT parameters. 
It will be important to 
test the stability of these 
results by including 
the analytic non-Gaussian covariance as in~\cite{Wadekar:2019rdu}, 
or using an empirical covariance from numerical simulations.

Our analysis suggests several directions for further improvements. 
From the simulation side, we have only analyzed the friends-of-friends halo catalogs,
which underestimate the non-linear redshift-space
distortions, known as the fingers-of-God~\cite{Jackson:2008yv}.
Thus, it will be important to extend our study to Rockstar~\cite{2013ApJ...762..109B} and COMPASO~\cite{Hadzhiyska:2021zbd}
halo finders, which captures fingers-of-God more accurately~(see e.g.~\cite{Schmittfull:2020trd,Ivanov:2024xgb} and references therein).
In addition, it will be interesting to apply our model 
to simulated galaxies whose properties are close to the realistic samples.

From the theory point of view, it will be interesting to extend our
model to higher order statistics, such as the $\La$-galaxy bispectrum. 
The particular configuration dependencies of higher order statistics
will allow one to break degeneracies between EFT parameters and eventually
improve cosmological constraints. Moving forward, it will be important to develop
simulation based priors for the $\La$ forest and high redshift galaxies
along the lines of~\cite{Ivanov:2024hgq,Ivanov:2024xgb}. 

The ultimate goal of EFT modeling is to infer 
cosmological parameters from the actual data. 
For that one has to extend our analysis to cosmological parameters,
starting with validations against mock data that resemble the actual 
observations in terms of clustering properties. The Sherwood data that we used here 
does not have large enough volume for this purpose, but it can be done with 
larger simulation suites such as ACCEL2~\cite{Chabanier:2024knr}.

Finally, it will be interesting to understand how the $\La$-quasar 
cross-correlations are affected in the presence of new physics.
For instance, it is known
that the bispectrum of different 
tracers can be used to test the equivalence principle~\cite{Peloso:2013zw,Kehagias:2013yd,Valageas_2011,Creminelli:2013mca,Creminelli:2013nua,Creminelli:2013poa}.
As example 
of such scenario is the violation of Lorentz invariance in the dark matter sector~\cite{Blas:2012vn,Audren:2014hza}. 
Since the $\La$ forest primarily tracers baryons (which obey the equivalence principle with high precision), while quasars trace dark matter, 
one could expect the $\La$-quasar bispectrum to be a sensitive probe 
of this scenario. 
We leave this and other research directions listed above for future exploration.

\vspace{1cm}
%%%%%%%%%%%%%%%%%%%%%%%%%%%%%%%%%%%%%%%
\section*{Acknowledgments}
% %%%%%%%%%%%%%%%%%%%%%%%%%%%%%%%%%%%%%%%%%%%%%%%%%%%%%%%%%%%%%%%
MI would like to thank Roger de Belsunce, Andrei Cuceu, Shi-Fan Chen and Martin White for 
enlightening conversations.
MI thanks Roger de Belsunce
for comments on the draft. 
AC acknowledges funding from the Swiss National Science Foundation.
Numerical calculations have been performed with the Helios cluster at the
Institute for Advanced Study, Princeton and the Baobab high-performance computing cluster at the University of Geneva.

\appendix 

\section{$P^\lya$ analysis}
\label{app:lya}

Here we present the results from the Ly$\alpha$ forest auto-power spectrum for various data cut choices.

\begin{figure*}
    \centering
    \includegraphics[width=0.99\textwidth]{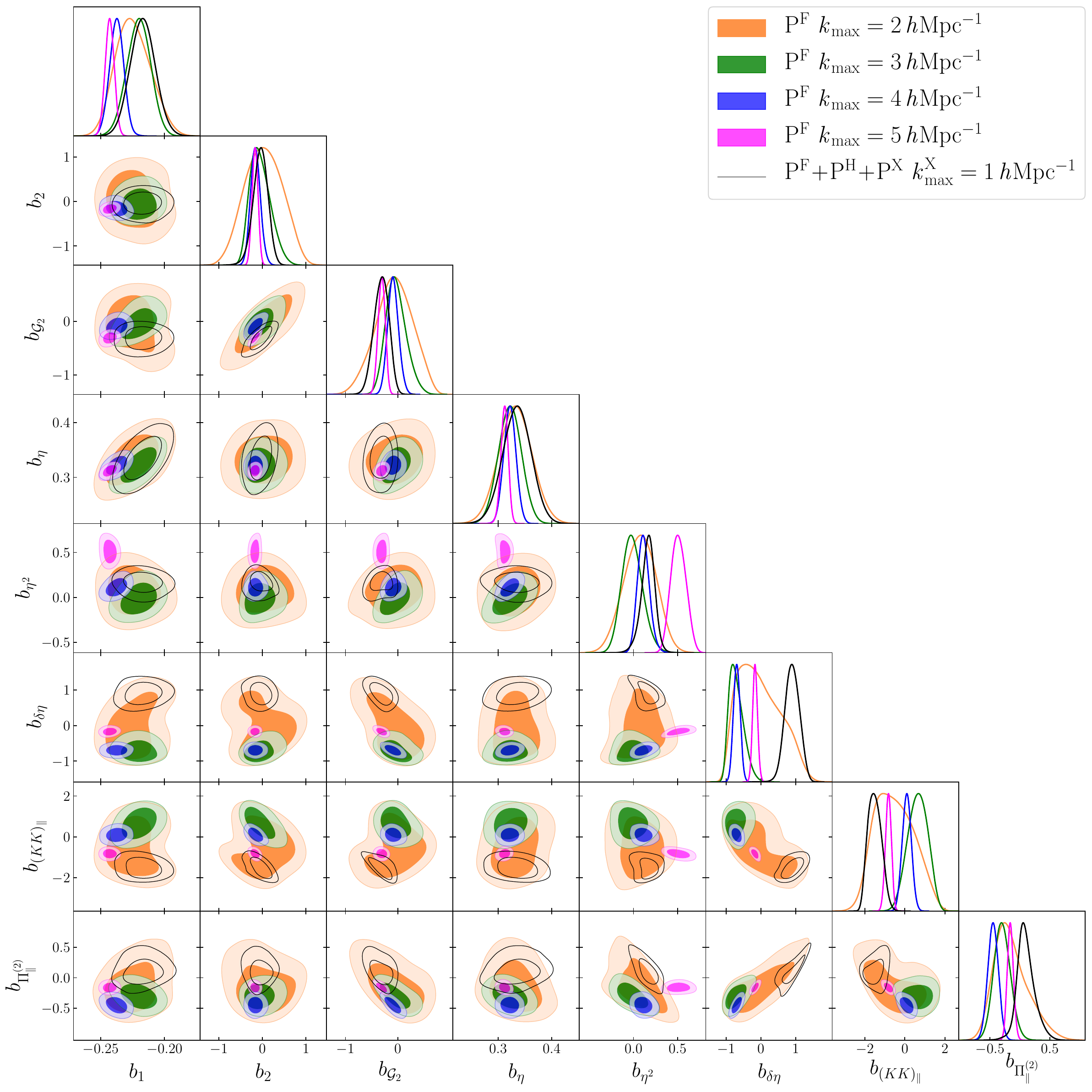}\\
    \caption{Marginalized posteriors on nuisance parameters of the EFT model for the Ly$\alpha$ forest auto-power spectrum $P^\lya$ for different values of $\kmax:$ $2$, $3$, $4$ and $5\hMpc$ (orange, green, blue, magenta, respectively). For comparison, the results from the baseline $P^\lya+P^\qso+P^\cross$ analysis, using $(\kmax^\lya,\kmax^\qso,\kmax^\cross)=(2,0.8,1)\hMpc$ with the diagonal covariance, are shown (black).}
    \label{fig:Lya}
\end{figure*}

Fig.~\ref{fig:Lya} illustrates the posterior distribution of the bias parameters.
Results are presented for four different $\kmax$ values: $2$, $3$, $4$, and $5\hMpc$. 
We see that the posteriors from the $P^\lya$-only analysis with $\kmax=2\hMpc$ are fully consistent with those of the baseline 3-spectra analysis. 
While the $\lyaa$ contours for $\kmax=3\hMpc$ are entirely consistent with $\kmax=2\hMpc$ results, they are inconsistent with the baseline 3-spectra analysis (which uses $\kmax^\lya=2\hMpc$).
This highlight the importance of multi-tracer analysis in validating scale cuts. 
For $\kmax=4\hMpc$ and $\kmax=5\hMpc$, the $P^\lya$-only analysis exhibits significant shifts and a dramatic reduction of the posterior volume.

\begin{figure}[!t]
    \centering
    \includegraphics[width=0.47\textwidth]{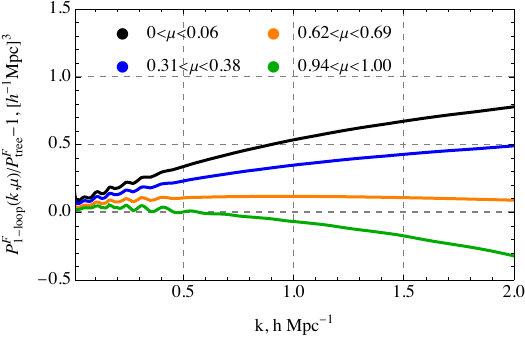}
    \caption{The magnitude of one-loop corrections relative to the linear theory prediction for the Ly$\alpha$ forest auto-power spectrum. The theory prediction is based on the best-fit model with $\kmax=2\hMpc$.}
    \label{fig:loop_lya}
\end{figure}

To evaluate the validity of the perturbation theory, we assess the magnitude of the one-loop correction as a function of wavenumber.
Fig.~\ref{fig:loop_lya} shows the one-loop
contribution divided by the tree-level model for the baseline configuration $\kmax=2\hMpc$. 
These results indicates that the perturbative approach is valid up to $\kmax=2\hMpc$.
Importantly, as shown in ref.~\cite{Ivanov:2023yla}, for $\kmax\sim 3\hMpc$, the one-loop corrections is comparable to the tree-level result, suggesting that higher loop corrections may be not negligible.
Fitting the $P^\lya$-only data up to $\kmax=3\hMpc$ leads to unphysical values of the EFT parameters, which are inconsistent with the results of the $P^\lya+P^\qso+P^\cross$ analysis.
The Ly$\alpha$ -- all-halo cross-power spectrum effectively break parameter degeneracies, ensuring unbiased inference of the EFT parameters.

\bibliographystyle{JHEP}
\bibliography{short.bib}

\end{document}